\documentclass[preprint,times,twocolumn,authoryear]{elsarticle}

\usepackage{amssymb}

\usepackage{amsmath}
\usepackage{amssymb}
\usepackage{epigraph}
\usepackage{graphicx}
\usepackage{tabularx}
\usepackage{epstopdf}
\usepackage{setspace}
\usepackage{dcolumn}
\usepackage{bm}
\usepackage{color}

\journal{Journal of Theoretical Biology}

\begin{document}

\parbox[t]{13.5cm}{This manuscript version is distributed under the CC-BY-NC-ND (Creative Commons) license.}
\vspace{2cm}

\parbox[t]{13.5cm}{The study has appeared as:\\
M{\"u}ller, E. J., van Albada, S. J., Kim, J. W., \& Robinson, P. A. (2017). Unified neural field theory of brain dynamics underlying oscillations in Parkinson’s disease and generalized epilepsies. \emph{Journal of Theoretical Biology}, 428, 132--146, DOI: 10.1016/j.jtbi.2017.06.016}

\newpage

\twocolumn[{
\begin{frontmatter}



\title{Unified neural field theory of brain dynamics in Parkinson's disease and generalized epilepsies}

\author[label1,label2]{E. J. M{\"u}ller \corref{cor1}}
\ead{eli.muller@sydney.edu.au}
\author[label1,label3]{S. J. van Albada}
\author[label1,label2]{J. W. Kim}
\author[label1,label2]{P. A. Robinson}

\cortext[cor1]{School of Physics, The University of Sydney, Sydney, 
NSW 2006, Australia}

\address[label1]{School of Physics, The University of Sydney, Sydney, 
NSW 2006, Australia}
\address[label2]{Center for Integrative Brain Function,
The University of Sydney NSW 2006, Australia}
\address[label3]{Institute of Neuroscience and Medicine (INM-6) and
Institute for Advanced Simulation (IAS-6) and JARA BRAIN Institute I,
J\"{u}lich Research Center, J\"{u}lich, Germany}

\begin{abstract}
The mechanisms underlying pathologically synchronized neural oscillations in Parkinson's disease~(PD) and generalized epilepsies are explored in parallel via a physiologically-based neural field model of the corticothalamic-basal ganglia~(CTBG) system.
%
The basal ganglia~(BG) are approximated as a single effective population and their roles in the modulation of oscillatory dynamics of the corticothalamic~(CT) system and vice versa are analyzed.
%
%
In addition to normal EEG rhythms, enhanced activity around~4~Hz and~20~Hz exists in the model, consistent with the characteristic frequencies observed in PD.
%
%
These rhythms result from resonances in loops formed between the BG and CT populations, analogous to those that underlie epileptic oscillations in a previous CT model, and which are still present in the combined CTBG system.
Dopamine depletion is argued to weaken the dampening of these loop resonances in PD, and network connections then explain the significant coherence observed between BG, thalamic, and cortical population activity around $4$--$8$~Hz and $20$~Hz.
%
%
Parallels between the afferent and efferent connection sites of the thalamic reticular nucleus~(TRN) and BG predicts low dopamine to correspond to a reduced likelihood of tonic-clonic~(grand mal) seizures, which agrees with experimental findings. Furthermore, the model predicts an increased likelihood of absence~(petit mal) seizure resulting from pathologically low dopamine levels; this agrees with experiments that show an increased likelihood of absence seizures following dopamine depletion.
%
%
Suppression of absence seizure activity is demonstrated when afferent and efferent BG connections to the CT system are strengthened, which is consistent with other CTBG modeling studies.
%
%
The BG are demonstrated as having a suppressive effect on activity of the CTBG system near tonic-clonic seizure states and suggests BG circuits as a target for treatment. 
%
%
Sleep states of the TRN are also found to suppress pathological PD activity in accordance with observations. 
%
%
Overall, the findings demonstrate strong parallels between coherent oscillations in generalized epilepsies and PD, and provide insights into possible comorbidities.
\end{abstract}


\end{frontmatter}
}]






\section{Introduction}
%
Parkinson's disease~(PD) and epilepsy are neurological disorders characterized by pathologically synchronous neural activity.
%
%
In generalized epilepsies, abnormal synchronization of population activity in
large assemblies occurs during a seizure and produces involuntary
paroxysmal alterations in behavior, including jerking movements,
transient loss of awareness, and in severe cases, massive convulsions and loss of consciousness \citep{Kandel91}. There are several forms of generalized epilepsy. However, we only consider two
of the most prominent seizure types: petit mal, and grand mal.
Petit mal (absence) seizures are defined by a brief period of unresponsiveness, identical pre- and post-seizure states, and prominent $\sim$3 Hz oscillations observed during adult human EEG recording \citep{Stefan97}. Grand mal (tonic-clonic) seizures consist
of an abrupt loss of consciousness and control of posture \citep{Kandel91}. Periods of increased muscle tone (tonic phase), displaying large $\sim$10 Hz oscillations, are followed by periods of jerking movement (clonic phase) with poly-spike wave complexes occupying the 1--4 Hz EEG frequency band.
%
%
%
%
In tremulous PD, enhanced oscillations are typically observed at~$4$--$8$~Hz and~$20$~Hz in activity of the basal ganglia~(BG)~\citep{Brown01,Tass10,Timmermann03,Wang05} and these seem to correlate with significant synchronization between the BG at these frequencies~\citep{Kuhn05,Levy02,Levy02dep,Marsden01,Rivlin06,Williams02}.
There is also a reported `slowing' of EEG signals in PD as the disease progresses with spectral power in the low frequency delta and theta bands increasing relative to the high frequency alpha and beta bands \citep{Morita09, Soikkeli91}.
Gradual degeneration of dopaminergic neurons in the substantia nigra pars compacta~(SNc) and ventral tegmental area~(VTA)~\citep{Bernheimer73,Ehringer60} leads to pathological changes in activity of BG and the neural circuits they are involved in.
It has been argued that PD oscillations are sustained in the circuit formed by the globus pallidus external segment (GPe) and the subthalamic nucleus (STN)~\citep{Humphries06,Plenz99,Terman02}.
%
%
However, incoming and outgoing connections of the BG parallel those of the thalamic reticular nucleus~(TRN), suggesting that PD rhythms could have similar dynamical properties to those of epilepsy.

%
%
A prominent interpretation of the BG is based on `direct' and `indirect' pathways through two partially segregated neural populations in the striatum, the main input station of the BG~\citep{Albin89,Alexander90,Parent95a}, as shown in Fig.~\ref{fig:Schem}(a). The striatum receives glutamatergic excitatory inputs from the cortex and specific relay nuclei~(SRN) of the thalamus and is divided into two populations, one expressing the D1 dopamine receptor and the other the D2 receptor~\citep{Jiang90}.
Inhibition projects onto the SRN via the globus pallidus internal segment~(GPi) and substantia nigra pars reticulata~(SNr).
The net effect of excitation incident to the BG on the levels of BG-induced inhibition of the SRN is dependent on the dominant internal BG pathway.
The direct pathway is argued to facilitate SRN disinhibition via a direct inhibitory coupling of the D1 striatal neurons to the GPi/SNr~\citep{Albin89,Parent95b,Smith98}.
The indirect pathway facilitates SRN inhibition whereby activity of the D2 striatal neurons results in excitation of the GPi/SNr via the GPe and STN. The hyperdirect pathway also facilitates inhibition of the SRN but via a direct coupling of cortical activity to the STN.
These three key pathways are shown in Fig.~\ref{fig:Schem}(a).
%
%
Dopamine appears to facilitate striatopallidal transmission via the direct pathway and inhibit transmission via the indirect pathway~\citep{Jankovic02,Kita11,Parent98}.
The reduced levels of dopamine observed in PD likely increase the overall responsiveness of the striatum to cortical excitation and strengthen transmission via the indirect pathway~\citep{Albin89,Delong90,Gerfen00,Mehler06}.

Physiologically based mean-field models of the brain allow tractable analysis of large-scale neuronal dynamics by averaging over microscopic structure~\citep{Freeman75,Jirsa96,Nunez74,Robinson97,Lopes74,Wilson73,Wright96}.
Neural field theory provides a mathematical framework for developing these models, which have been successful in accounting for many characteristic states of brain activity including sleep stages, eyes-open, and eyes-closed in wake, nonlinear seizure dynamics and many other phenomena~\citep{Breakspear06,Jirsa96,Liley05,Roberts08,Robinson02,Robinson97,Robinson98,Steyn04}. The theory incorporates realistic anatomy of neural populations, non-linear neural responses, interpopulation connections and dendritic, synaptic, cell-body, and axonal dynamics~\citep{Deco08,Rennie99,Rennie00,Robinson05,Robinson97,Robinson01,Robinson98,Wilson73,Wright96}.
Significant progress has been made in the modeling of epilepsy \citep{Breakspear06, Robinson02, Robinson01, Wendling01, Wendling00}. A corticothalamic (CT) model is able to account
for the nonlinear features observed in EEGs of generalized epilepsy
as well as seizure activation by stimuli \citep{Breakspear06, Roberts08, Robinson02, Robinson01} and has provided insights into the genesis of pathological rhythms.
%
%

%
\begin{table}
\caption{\label{tab:abbr} Abbreviations used.}
\begin{tabularx}{\columnwidth}{ l l}
\hline 
Term & Abbreviation\\ 
\hline
Parkinson's disease & PD\\
Basal ganglia & BG \\
Specific relay nuclei & SRN \\
Thalamic reticular nucleus & TRN \\
Globus pallidus internal segment & GPi \\
Globus pallidus external segment & GPe \\
Subthalamic nucleus & STN \\
Corticothalamic & CT \\
Corticothalamic-basal ganglia & CTBG \\
\hline
\end{tabularx}
\end{table}
%
%
%
%
\cite{Albada09a} and~\cite{Albada09b} developed a neural field theory of the corticothalamic-basal ganglia~(CTBG) system that accounted semiquantitatively for several key electrophysiological correlates of PD. Significantly, the model demonstrated changes in average activity of various populations, $\sim$$5$~Hz and $\sim$$20$~Hz oscillations, altered responses to transient cortical stimulation, and changes in EEG spectra.
%
%
There have also been several recent modeling studies aimed at describing the role of the BG in modulating seizure activity~\citep{Chen14,Chen15,Hu15,Hu15dbs}. Suppression of absence seizure oscillations was achieved by increasing SNr activity which is coupled to the thalamic populations~\citep{Chen14,Hu15} and by strengthening GPe inhibitory coupling to the cortex~\citep{Chen15}. Another study used a CTBG model to demonstrate the control of absence seizure activity by deep brain stimulation~(DBS) of the SNr and the cortex~\citep{Hu15dbs}.
Although these models have progressed our understanding of PD and seizure dynamics in the CTGB system, their complexity means that it is difficult to obtain insights from them.

%
%
%
The core aim of this paper is thus to develop and explore a simplified unified neural field model of the CTBG system that conserves features essential to the production of epileptic and parkinsonian states and provides insight into their dynamics and interactions, especially their characteristic frequencies.
%
%
%
The unified framework will allow us to explore possible analogs between the two disorders,  interpret the synchronous activity observed between BG nuclei in PD, and will provide insight into the modulation of seizure activity by the BG and of PD by the TRN.

This work makes extensive use of abbreviations and these have been summarized in Table~\ref{tab:abbr}.
%
%

%
In Sec.~\ref{sec:nf} we outline the neural field theory of our simplified CTBG model. We give the methods used in computing steady states and stability analysis, and the derivation of the system's transfer function.
In Sec.~\ref{sec:symm} we exploit symmetries in the connectivity of the BG and TRN populations. 
We use parameter values describing the BG that are equal to those used to describe the TRN. We outline BG modulation of seizure states and how these interrelate with parkinsonian states. 
In Sec.~\ref{sec:asymm} we analyze more general configurations of the model using parameters drawn from the literature. We analyze key resonances in the system that underlie global oscillations, their modulation by the BG, and the nature of parkinsonian states. We also investigate BG modulation of absence seizure activity and tonic-clonic activity and possible modulation of parkinsonian activity by the TRN.
In Sec.~\ref{sec:disc} we summarize our findings and discuss their implications.

\section{Neural Field Theory}
\label{sec:nf}
In this section a summary of neural field theory is given~\citep{Beurle56,Wilson72,Wilson73,Nunez74,Jirsa96,Freeman75,Robinson97,Rennie99,Albada09a} and how it is applied to develop a simplified CTBG model.


\subsection{Simplified model}
\label{subsec:simp}
\begin{figure*}[t]
\centering
\includegraphics[width=0.8\textwidth]{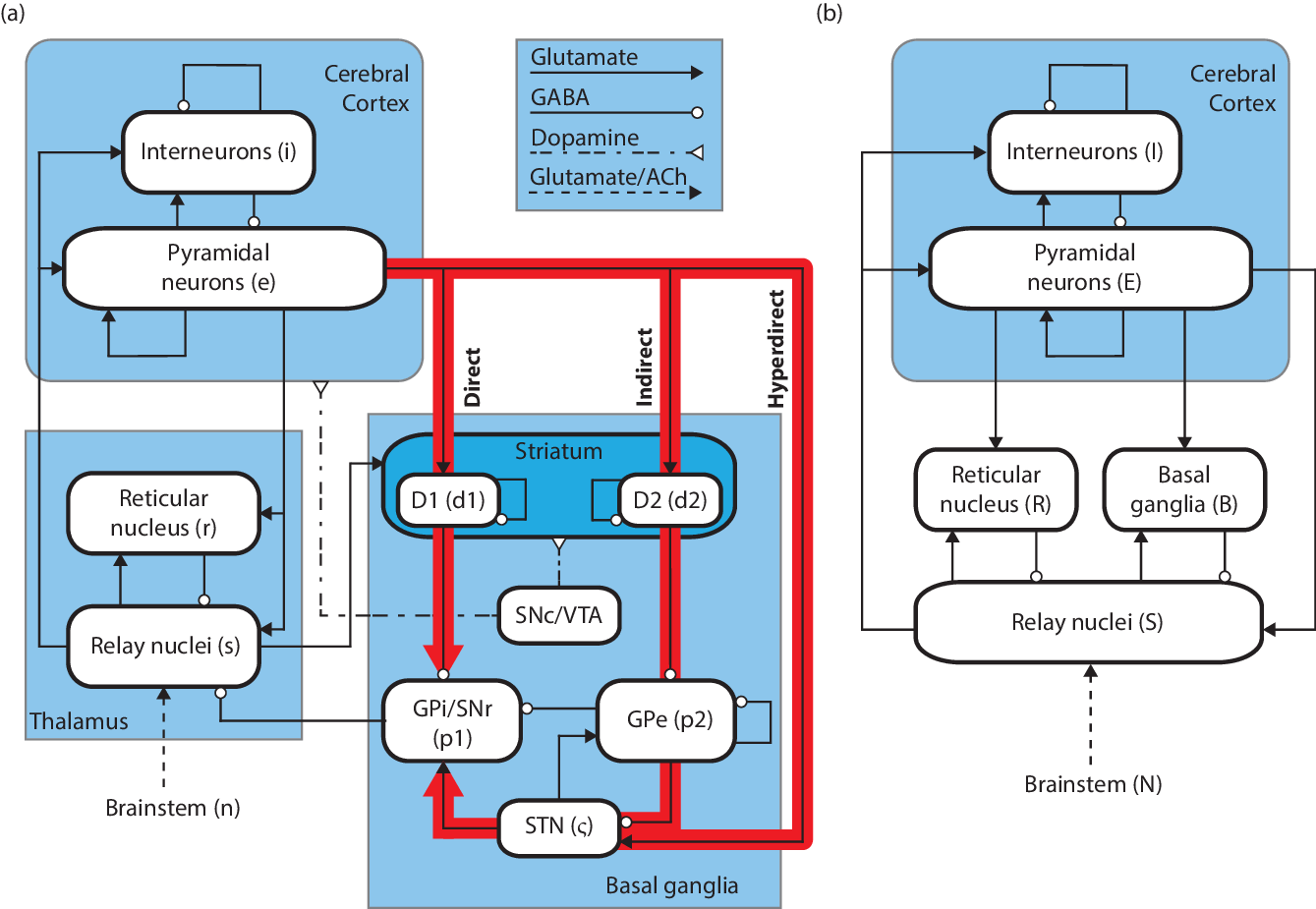}
\caption{\label{fig:Schem} Schematics of the corticothalamic-basal ganglia system. Subscripts used to denote various neural populations are parenthesized. Arrowed lines denote neural connections and corresponding neurotransmitters, glutamate, GABA, dopamine, and acetylcholine (ACh). (a) CTBG model of \cite{Albada09a} describing the BG using five populations: D1 and D2 are striatal neurons expressing dopamine receptors of the corresponding classes; GPi, GPe are globus pallidus internal and external segments; SNr, SNc are substantia nigra pars reticulata and pars compacta; STN is the subthalamic nucleus. (b) Simplified CTBG model proposed in this paper. The BG are approximated as a single effective population whose connections parallel those of the TRN.}
\end{figure*}

The core aim of the present work is to simplify previous approaches to modeling the CTBG system.
Van Albada and Robinson (2009) and \cite{Albada09b}, using a more detailed model, observed the indirect pathway through the BG to be dominant for physiological parameter ranges. This suggests that the full BG system, shown in Fig.~\ref{fig:Schem}(a), can be effectively approximated as a single population.
We thus propose a simplified BG population that facilitates dominant indirect pathway dynamics whereby afferent BG excitation inhibits the SRN.
Activity of this population is considered to represent GPi/SNr activity as these populations project inhibition monosynaptically onto the SRN.
As shown in Fig~\ref{fig:Schem}(b) the connectivity of the simplified BG approximates that of the full system.

The simplified CTBG model contains five neural populations.
The cortex is considered to contain populations of excitatory pyramidal neurons, $E$, and inhibitory interneurons, $I$. The thalamus is divided into the SRN, $S$, which are considered as purely excitatory and the TRN, $R$, which are inhibitory. As mentioned above, the BG are approximated as a single population, $B$, whose efferent activity inhibits the SRN analogously to the TRN's effect.

\subsection{Firing rates}
The mean firing rate, $Q_a(t)$, of a population can be approximately related to its mean membrane potential, $V_a(t)$, \citep{Wilson72} by
\begin{align}
\label{eq:sig}
Q_a (t) & = S_a[V_a(t)], \\
	& = \frac{Q_a^{\rm max}}{1+{\rm exp}[-\{V_a(t)-\theta_a\}/\sigma^\prime]},
\end{align}
where $a= E,I,R,S$, or $B$. This sigmoidal response function saturates at the maximal firing rate $Q_a^{\rm max}$, $V_a$ is the average membrane potential relative to resting, $\theta_a$ is the mean neural firing threshold, and $\sigma^\prime \pi /\sqrt{3}$ is the standard deviation of this threshold.


\subsection{Axonal propagation}
A number of experimental studies have revealed waves of neural activity spreading across the cortex~\citep{Burns51,Nunez74,Chervin88,Golomb97}, which have been analyzed theoretically~\citep{Beurle56,Bressloff01,Bressloff03,Jirsa96,Jirsa97,Nunez95,Robinson97,Deco08}.
This propagating activity is represented as a field of mean spike rates in axons, $\phi_a$. A population $a$ with a mean firing rate $Q_a$ that is related to the field by the damped wave equation,
\begin{equation}
\label{eq:field_spike}
     D_a(t)\phi_a(t) = Q_a (t),
\end{equation}
where
\begin{equation}
\label{eq:diff_damp_wave}
     D_a(t) = \frac{1}{\gamma_a^2} \frac{\partial^2}{\partial t^2} +
     \frac{2}{\gamma_a} \frac{\partial}{\partial t} + 1.
\end{equation}
Here, $\gamma_a = v_a/r_a$ represents the damping rate, where $v_a$ is the propagation velocity in axons, and $r_a$ is the characteristic axonal length for the population. The propagation of these waves is facilitated primarily by the relatively long-range white matter axons of excitatory cortical pyramidal neurons.
Later in our model the simplifying local interaction approximation $r_b \approx 0$ is made for $b = I, R, B, S$ due to the short ranges of the corresponding axons which implies $\phi_b({\bf r}, t) = Q_b({\bf r}, t)$ for these populations~\citep{Robinson97,Rennie99,Robinson01,Robinson02,Robinson04,Rowe04}.


\subsection{Synaptodendritic response}
Afferent spike activity at the dendritic tree is integrated at the soma, causing time-dependent perturbations from the resting potential. 
The influence of incoming spikes to population $a$ from population $b$ is weighted by a connection strength parameter, $\nu_{ab} = N_{ab}s_{ab}$, where $N_{ab}$ is the mean number of connections between the two populations and $s_{ab}$ is the mean strength of response in neuron $a$ to a single spike from neuron $b$. The soma potential is thus modeled as
\begin{equation}
\label{eq:soma_response}
D_{\alpha \beta} V_a (t)  = 
      \sum_b \nu_{ab} \phi_b (t - \tau_{ab}),
\end{equation}
with
\begin{equation}
\label{eq:dab}
D_{\alpha \beta} = 
     \frac{1}{\alpha_a \beta_a}\frac{d^2}{dt^2} + 
       \left(\frac{1}{\alpha_a} + \frac{1}{\beta_a} \right) \frac{d}{dt} + 1. 
\end{equation}

Here, $\tau_{ab}$ is the average axonal delay for the transmission of signals to population $a$ from population $b$. The differential operator $D_{\alpha \beta}$ gives the soma potential response allowing for synaptodendritic dynamics and soma capacitance; it has the effect of low-pass filtering the inputs~\citep{Rennie00,Robinson97}. In this work we use an effective response function for each population, which assumes equal response rates for both glutamatergic excitation and GABAergic inhibtion.



\subsection{Random connectivity approximation}
Following \cite{Wright96}, \cite{Robinson98} and \cite{Braitenberg98}, symmetries in the intracortical connection strengths are exploited. Connections in the cortex are approximated as random and thus the number of each type of synapse is proportional to the number of neurons; i.e., if $80\%$ of the neurons in the cortex are excitatory then $80\%$ of pre- and post-synaptic connections will be with excitatory neurons. This approximation results in $\nu_{EE} = \nu_{IE}$,  $\nu_{EI} = \nu_{II}$ and $\nu_{ES} = \nu_{IS}$. If excitatory and inhibitory populations are assumed to have equal response function,~(\ref{eq:dab}), this implies $V_E = V_I$ and $Q_E = Q_I$. Inhibitory population variables can then be expressed in terms of excitatory quantities as a further simplification and thus are not neglected even though they do not appear explicitly below.


\subsection{Steady states}
\label{subsec:steady_states}
The steady-state firing rates of the model are of particular interest and relate directly to experiments. Spatially uniform steady states are obtained by setting all time derivatives to zero in Eqs~(\ref{eq:field_spike})--(\ref{eq:dab}). With the connectivity shown in Fig.~\ref{fig:Schem}(b), Eq.~(\ref{eq:soma_response}) then becomes
\begin{align}
	\label{eq:soma_Ve} V_E^{(0)} &= \left(\nu_{EE} + \nu_{EI}\right)\phi_{E}^{(0)} + \nu_{ES}\phi_{S}^{(0)},\\
	\label{eq:soma_Vr}V_R^{(0)} &= \nu_{RE}\phi_{E}^{(0)} + \nu_{RS}\phi_{S}^{(0)},\\
	\label{eq:soma_Vb}V_B^{(0)} &= \nu_{BE}\phi_{E}^{(0)} + \nu_{BS}\phi_{S}^{(0)},\\
	\label{eq:soma_Vs}V_S^{(0)} &= \nu_{SE}\phi_{E}^{(0)} + \nu_{SR}\phi_{R}^{(0)} + \nu_{SB}\phi_{B}^{(0)} + \nu_{SN}\phi_N^{(0)},
\end{align}
where the superscripts (0) denote steady-state values.

Following the approach of \cite{Robinson98,Robinson04}, elimination of other variables allows a single transcendental equation to be obtained for $\phi_{E}^{(0)}$ as a function of $\phi_N^{(0)}$:
\begin{equation}
\label{eq:fp}
\begin{split}
& S_S^{-1} \left\{\frac{1}{\nu_{ES}} \left[S_E^{-1}\left(\phi_E^{(0)}\right)
- (\nu_{EE}+\nu_{EI})\phi_E^{(0)}\right] \right\} \\
& = \nu_{SE}\phi_E^{(0)} + \nu_{SN}\phi_N^{(0)} \\
& + \nu_{SB} S_B \left\{ \nu_{BE}\phi_E^{(0)} + \frac{\nu_{BS}}{\nu_{ES}}
\left[ S_E^{-1} \left(\phi_E^{(0)} \right) - (\nu_{EE}+\nu_{EI})\phi_E^{(0)}\right] \right\} \\
& + \nu_{SR} S_R \left\{ \nu_{RE}\phi_E^{(0)} + \frac{\nu_{RS}}{\nu_{ES}}
\left[ S_E^{-1} \left(\phi_E^{(0)} \right) - (\nu_{EE}+\nu_{EI})\phi_E^{(0)}\right] \right\},
\end{split}
\end{equation}
where $\phi_b^{(0)} = Q_b^{(0)}$ in the steady state and the function $S_a^{-1}$ is the inverse of the sigmoid Eq.~(\ref{eq:sig}), with
\begin{equation}
\label{eq:inv_sig}
S_a^{-1}(\phi_a) = V_a = \theta_a - \sigma'\ln \left(\frac{Q_a^{\rm max} - \phi_a}{\phi_a}\right).
\end{equation}
The excitatory firing rate, $\phi_E^{(0)}$ can be found by solving Eq.~(\ref{eq:fp}) using a numerical bisection method for a low intensity white noise input, $\phi_N^{(0)}$. The remaining populations' steady-state firing rates can be calculated using Eqs~(\ref{eq:soma_Ve})-(\ref{eq:soma_Vs}). In general there is an odd number of steady-states~\citep{Robinson98,Robinson04}.
The single low firing solution corresponds to a normal state. Other steady-state solutions are
typically either unstable or have extremely high firing rates and represent physiologically unrealistic states~\citep{Noroozbabaee16,Robinson97,Robinson98,Robinson02,Robinson04}. 


\subsection{Transfer function}
\label{subsec:transfer}
In this section the CTBG system's transfer function is derived which is later used to generate the power spectra of population firing rates in response to white noise input.
One of the main advantages of using a mean-field description of the brain is that population activity can easily be related to electroencephalographic (EEG) scalp recordings and local field potential~(LFP) measurements, which are themselves averages.
EEG and LFP signals result from fields generated by the movement of charge in and out of neurons during firing and signal propagation.
Charge flow occurs in closed loops, causing extracellular potentials that can be approximated as dipoles~\citep{Nunez95,Nunez06}.
Thus, fluctuations in population activity $\phi_a$ are related most directly to voltage fluctuations measured during LFP recording and EEG~\citep{Srinivasan96,Robinson01,Nunez06}.

%
%
To derive the CTBG system's transfer function, perturbations relative to the steady-state, $V^{(0)},Q^{(0)}=S(V^{(0)})$, are considered small enough that nonlinear effects can be ignored. A Taylor expansion of the sigmoidal function~(\ref{eq:sig}) is made about the steady-state from Sec.~\ref{subsec:steady_states} with
\begin{equation}
\label{eq:tay_exp}
Q_a(t) - Q_a^{(0)} = \left.\frac{dQ_a}{dV_a}\right|_{V_a^{(0)}}\left[V_a(t) - V_a^{(0)}\right] + ...
\end{equation}
Taking only first-order terms denoted by the superscript $(1)$, Eq.~(\ref{eq:tay_exp}) can be expressed as
\begin{equation}
\label{eq:lin_app}
Q_a^{(1)}(t)\approx \rho_a V_a^{(1)}(t),
\end{equation}
where
\begin{equation}
\label{eq:sig_diff}
\rho_a = \left.\frac{dQ_a}{dV_a}\right|_{V_a^{(0)}} = \frac{\phi_a^{(0)}}{\sigma'}\left[1-\frac{\phi_a^{(0)}}{Q_a^{\rm max}}\right]
\end{equation}
and $\phi_a^{(0)} = Q_a^{(0)}$ in the steady-state limit. 
To simplify the equations the superscripts on $Q_a^{(1)}, \phi_a^{(1)}, V_a^{(1)}$ are dropped henceforth, i.e., $Q_a^{(1)}$ is written as $Q_a$.
The damped wave equation (\ref{eq:field_spike}) is then approximated as,
\begin{equation}
\label{eq:linear_field}
D_a(t)\phi_a(t) = Q_a(t) \approx \rho_a V_a.
\end{equation}

Substituting Eq.~(\ref{eq:linear_field}) into Eq.~(\ref{eq:soma_response}) gives
\begin{align}
\label{eq:field_set}
& D_{\alpha\beta}(t)D_{E}(t)\phi_E(t) = \left(G_{EE} + G_{EI}\right)\phi_{E}(t)\nonumber\\
	& \qquad +  G_{ES}\phi_{S}(t - \tau_{ES}),\\
& D_{\alpha\beta}(t)\phi_R(t) = G_{RE}\phi_{E}(t - \tau_{RE}) + G_{RS}\phi_{S}(t - \tau_{RS}),\\
& D_{\alpha\beta}(t)\phi_B(t) = G_{BE}\phi_{E}(t - \tau_{BE}) + G_{BS}\phi_{S}(t - \tau_{BS}),\\
& D_{\alpha\beta}(t)\phi_S(t) = G_{SE}\phi_{E}(t - \tau_{SE}) + G_{SR}\phi_{R}(t - \tau_{SR})\nonumber\\
	& \qquad + G_{SB}\phi_{B}(t - \tau_{SB}) + G_{SN}\phi_N(t),
\end{align}
where $G_{ab} = \rho_{a}\nu_{ab}$ is an average connection gain representing the additional activity in population $a$ per additional unit input activity from population $b$.
Upon transforming Eq.~(\ref{eq:field_set}) into the Fourier domain, $\phi_{E}(\omega)$ can be expressed in terms of $\phi_{N}(\omega)$, to obtain the transfer function
\begin{equation}
\label{eq:trn}
\frac{\phi_E(\omega)}{\phi_N(\omega)} =  \frac{1}{q^2r_e^2}
    \frac{J_{ES}J_{SN}}{(1-J_{EI})(1-J_{SRS}-J_{SBS})},
\end{equation}
where
\begin{align}
\label{eq:qq2}
q^2r_e^2  & =  \left(1-\frac{i \omega}{\gamma_E} \right)^2\nonumber \\
     &- \frac{1}{1-J_{EI}} 
     \left[ J_{EE} + \frac{J_{ESE}+J_{ESRE}+J_{ESBE}}{1-J_{SRS}-J_{SBS}}
     \right],
\end{align}
with
\begin{eqnarray}
L_a & = & \left(1-i\frac{\omega}{\alpha_a}\right)^{-1}
          \left(1-i\frac{\omega}{\beta_a}\right)^{-1},\\
J_{a_1 a_2} & = & L_{a_1} G_{a_1 a_2} \text{exp}\left(i\omega \tau_{a_1 a_2}\right),\label{eq:Ja1a2}\\
J_{a_1 a_2 \cdots a_n} & = &  J_{a_1 a_2}J_{a_2 a_3} \cdots J_{a_{n-1} a_n}.
\end{eqnarray}
Here, $J_{a_1 a_2 \cdots a_n}$ represents both the connection strengths and time delays associated with particular loops in Fig.~\ref{fig:Schem}(b); $J_{SRS}$ corresponds to the loop between SRN and
TRN, $J_{SBS}$ to the loop between BG and SRN, $J_{ESE}$ and $J_{ESRE}$ to the direct and
indirect CT loops, respectively, and $J_{ESBE}$ to the cortico-BG-SRN loop. In the case that
there is no BG population, $J_{SBS}=0$ and $J_{BSBE}=0$, and the transfer function (\ref{eq:trn}) reduces to that of the purely CT system~\citep{Robinson02}.

The power spectrum $P(\omega)$ is given by
\begin{equation}
\label{eq:powerspec}
P(\omega)= |\phi_E(\omega)|^2.
\end{equation}
%


\subsection{Stability analysis}
\label{subsec:stab_a}
As in \cite{Robinson01}, linear waves at $k=0$ in our model are described
by poles of the transfer function, Eq.~(\ref{eq:trn}), with
\begin{equation}
\label{eq:disp}
(1 - J_{EI})(1 - J_{SRS} - J_{SBS})q^2r^2_{E} = 0.
\end{equation}
%
%
%
Roots of Eq.~(\ref{eq:disp}) give dispersion relations for all possible
modes of the system, where $\omega$ is complex in general. The steady-state
solutions found in Sec.~\ref{subsec:steady_states} are linearly stable when all modes
satisfy Im~$\omega < 0$. The onsets of linear instabilities correspond to roots
of Eq.~(\ref{eq:disp}) where $\omega$ is purely real~\citep{Robinson97,Robinson02,Roberts12}.
%


\subsection{Repartitioning of CT parameters in the CTBG model}
\label{subsec:repart}
Neural field models of the CT system alone did not define any separate population(s) to represent the BG~\citep{Robinson01,Robinson02,Robinson04}.
This means that the effects of the BG were implicitly incorporated into the parametrization of the thalamus; i.e., thalamic populations used in the CT model implicitly represented the effects of both the thalamus and the BG.
The parameters for the exclusively CT system used to describe the TRN must thus be partitioned between the TRN and BG when they are explicitly kept separate.
%
%
Van Albada and Robinson (2009) and~\cite{Albada09b} performed this repartitioning for a more detailed CTBG model.
In the simplified CTBG model presented here, we similarly divide the TRN parameters from the exclusively CT system between the TRN and the BG.

\section{Symmetric CTBG system and reduced parameter space}
\label{sec:symm}
In our simplified CTBG system, afferent and efferent connections of the BG population parallel
those of the TRN, as seen in Fig.~\ref{fig:Schem}. The system dynamics are first explored in the most tractable form by approximating parameters of the BG and TRN populations as being identical. The connection strengths, axonal time delays, soma response rates, and firing rate maximums for the two populations are set to be equal and are given in Table~\ref{tab:symm_params}. 
The symmetric configuration allows us to represent the model in a dimensionally reduced form and tractably explore the effects of additional loop gains involving the BG.
In Sec.~\ref{sec:asymm} the more general asymmetric case is considered using parameter estimates taken from the literature and drawing on the findings of the present section.

\begin{table}
\caption{\label{tab:symm_params} Nominal parameters for the symmetric system, adapted
from \cite{Robinson02}.}
\begin{tabularx}{\columnwidth}{ l X X r }
\hline 
Quantity & & Value & Unit\\ 
\hline
$r$ 						&	& $80$ 		& mm			\\
$\sigma\prime$ 					&	& $3.3$ 	& mV			\\
$\phi_{N}^{(0)}$				&	& $1$ 		& $\text{s}^{-1}$	\\
$\alpha_a$					&	& $50$ 		& $\text{s}^{-1}$	\\
$\beta_a$					&	& $200$ 	& $\text{s}^{-1}$	\\
$t_0$						&	& $80$ 		& ms			\\
$\gamma_{E}$					&	& $116$ 	& $\text{s}^{-1}$	\\
$Q_a^{\rm max}$ 				&	& $250$ 	& $\text{s}^{-1}$	\\
$\theta_a$ 					&	& $15$ 		& mV			\\
$\nu_{EE}$					&	& $1.2$ 	& mV s			\\
$\nu_{EI}$ 					&	& $-1.8$ 	& mV s			\\
$\nu_{ES}$ 					&	& $1.2$ 	& mV s			\\
$\nu_{RE}$, $\nu_{BE}$ 				&	& $0.4$ 	& mV s			\\
$\nu_{RS}$, $\nu_{BS}$ 				&	& $0.2$ 	& mV s			\\
$\nu_{SE}$ 					&	& $1.2$ 	& mV s			\\
$\nu_{SR}$, $\nu_{SB}$ 				&	& $-0.8$ 	& mV s			\\
$\nu_{SN}$ 					&	& $0.5$ 	& mV s			\\
\hline
\end{tabularx}
\end{table}
%


\begin{figure*}[t]
\centering
\includegraphics[width=0.55\textwidth]{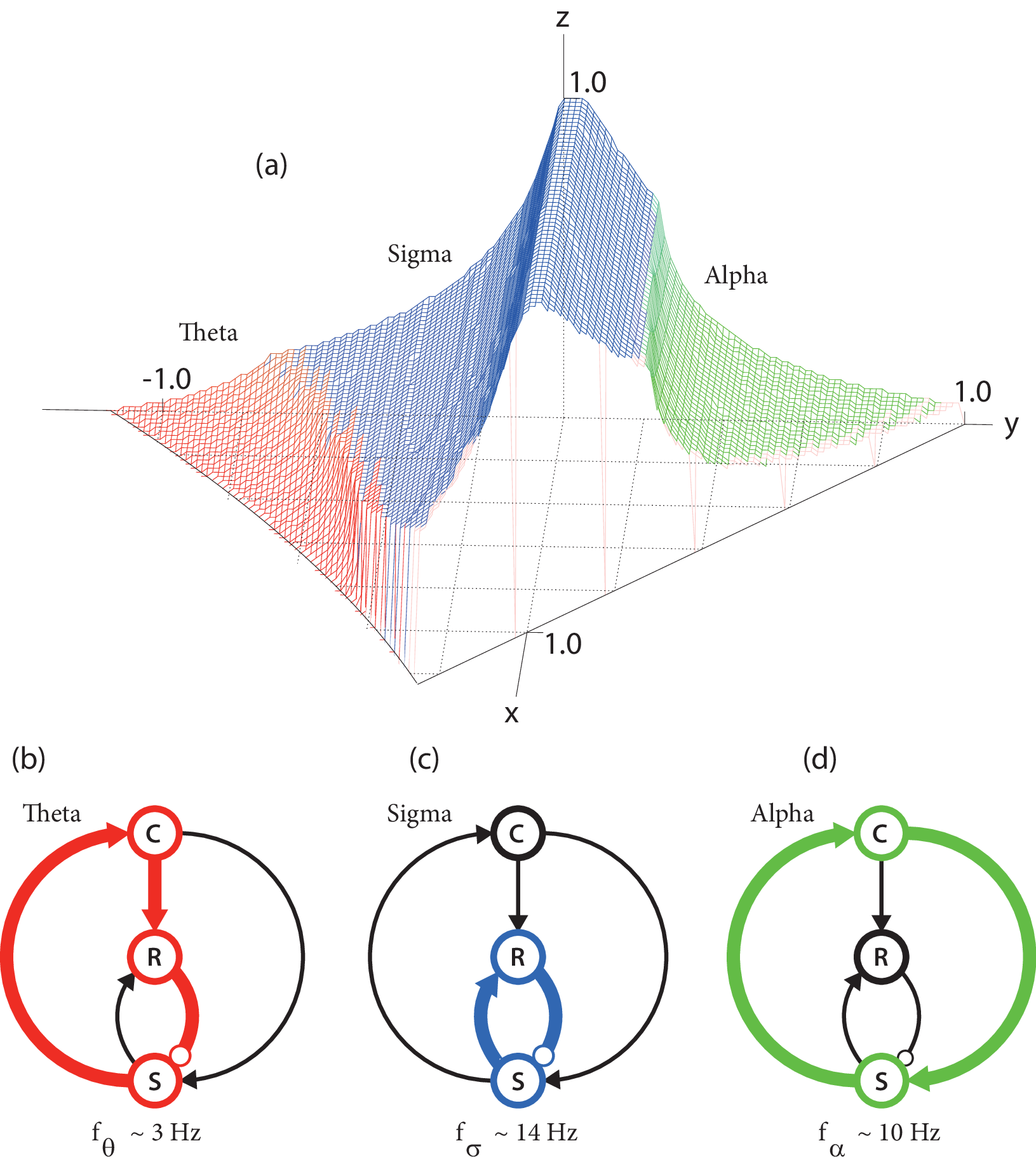}
\caption{\label{fig:ct_sch} Visualization of stability in the corticothalamic system and
a schematic comparison of three key neural circuits and their intrinsic periodicity.
(a) Linear stability zone of the corticothalamic model ~\citep{Robinson02}. States of the model within the dimensionally reduced space are represented by stability parameters $x$ for intracortical, $y$ for corticothalamic-BG, and $z$ for intrathalamic and thalamo-BG loops driving key instabilities. Stable points lie inside the surface and the onset of nonlinear oscillations occurs as the system passes outside the stability zone. The shaded regions define parameter values at which the system loses stability at delta/theta~(red), spindle/sigma~(blue), and alpha~(red) frequencies. (b) Delta/theta resonance: \textit{ESRE} loop generating $\sim$$3$~Hz oscillations. (c) Sigma resonance: \textit{SRS} loop generating $\sim$$14$~Hz spindle oscillations. (d) Alpha resonance: \textit{ESE} loop generating $\sim$$10$~Hz oscillations. E,~I: cortex, S: specific relay nuclei~(SRN), R: thalamic reticular nucleus~(TRN).}
\end{figure*}
%

\label{subsec:reduced_sys}
In the CT model of \cite{Robinson02}, the system dynamics are largely determined by three key loops, which allows for the approximate representation of the CT model's state and stability in a three-dimensional space.
This approximation provides a simplified means of visualizing and understanding key instability boundaries and highlights regions of parameter space where specific resonances produce characteristic spectral features.
Here, the approach is further generalized to incorporate the symmetrically configured BG population, prior to allowing for asymmetries in Sec.~\ref{sec:asymm}.
%
%
The only nonzero axonal delays $\tau_{ab}$ are $\tau_{SE} = \tau_{RE} = \tau_{BE}$, and $\tau_{ES}$ owing to the anatomical separation of the cortex, BG, and thalamus.
%
%
These nonzero delays can be consolidated in the transfer function into a single effective delay, $t_0$, given in Table~\ref{tab:symm_params}, that represents both the axonal delay in the
corticothalamic loop and the cortico-BG-thalamic loop, which are equal in the present symmetric case.

To simplify the analysis, the approximation $L_a = L \approx 1$ in Eq.~(\ref{eq:disp}) is made in the low frequency limit $\omega \ll \alpha, \beta, \gamma$ \citep{Robinson02} except in the $1 - L^2 (G_{SRS} + G_{SBS})$ term where the effect of $L$ must be retained to reproduce spindle oscillations. The low frequency approximation is only implemented in this section
to allow for a simplified representation of stability boundaries and
not in the later sections of this work. The approximation enhances
higher frequency terms in the systems transfer function that would
otherwise be more strongly low-pass filtered. However, it has been
verified to make only moderate quantitative differences to the stability of the system at low frequencies without changing key properties \citep{Robinson02}. By substituting Eq.~(\ref{eq:qq2}) into Eq.~(\ref{eq:disp}) the dispersion relation can then be expressed as
\begin{equation}
\label{eq:q2_0}
\left(1-\frac{i \omega}{\gamma_E} \right)^2 - x - y 
     \frac{1+(z/z_s)}{1+L^2 (z/z_s)} e^{i\omega t_0} = 0,\\
\end{equation}
where
\begin{eqnarray}
\label{eq:x}
x & = & \frac{G_{EE}}{1-G_{EI}},\\
\label{eq:y}
y & = &\frac{G_{ESE}+G_{ESRE}+G_{ESBE}}{(1-G_{EI})(1-G_{SRS}-G_{SBS})},\\
\label{eq:z}
z & = & (G_{SRS}+G_{SBS}) z_s,\\ 
z_s & = & - \frac{\alpha\beta}{(\alpha+\beta)^2}.
\end{eqnarray}
Here, $x$ parametrizes the total gain of the intracortical loops, $y$ parametrizes the net gain of the corticothalamic-basal ganglia loops, and $z$ parametrizes the gains of the intrathalamic loop and the thalamo-BG loop. Together these parameters determine the stability of the whole CTBG system.
The signs of the excitatory and inhibitory gains are physiologically determined and thus $x\geq 0$ and $z\geq 0$.
The reduced dimensionality resulting from this parametrization allows a surface to be defined
that bounds a zone of stable steady states, as shown in Fig.~\ref{fig:ct_sch}(a).
The stability boundary is given by solutions of Eq.~(\ref{eq:q2_0}) for real $\omega$.
Since connections involving the BG exactly parallel those involving the TRN in the symmetric CTBG system, the stability zone for the CT model~\citep{Robinson02} is further generalized by repartitioning the previous TRN gains equally between the TRN and BG in the symmetric CTBG system, as described in Sec.~\ref{subsec:repart}.

\cite{Robinson02} demonstrated that normal states of the brain lie within the stability zone 
of the CT model as seen in Fig.~\ref{fig:ct_sch}(a). Epileptic states lie at or beyond the
boundary of this zone, where the fixed point loses stability, and are characterized by highly ordered, large amplitude limit cycle oscillations~\citep{Robinson02,Breakspear06}.
The stability boundary is then divided into areas corresponding to the frequency bands these oscillations occupy and the dominant neural circuits responsible for their generation~\citep{Robinson02,Breakspear06}:
%
%

%
(i)~The left boundary in Fig.~\ref{fig:ct_sch}(a) corresponds to oscillations, typically at 2--5~Hz, and represents the transition to absence seizure states of the model~\citep{Robinson02,Breakspear06}.
In this region the parameter $y$ is negative with $|G_{ESE}| < |G_{ESRE} + G_{ESBE}|$ and $G_{ESRE} = G_{ESBE} < 0$. We refer to this boundary as the theta instability boundary, although it also includes most of the traditional delta frequency band.
%
%
(ii)~The right boundary corresponds to large amplitude oscillations with a peak frequency observed in the alpha band ($\sim$$10$ Hz) and represents a transition to tonic-clonic seizure states in the model~\citep{Robinson02,Breakspear06}. For this region of parameter space $y$ is positive because $G_{ESE} > 0$ and $|G_{ESE}| > |G_{ESRE} + G_{ESBE}|$. This boundary is referred to as the alpha instability boundary.
%
%
(iii)~The upper boundary of Fig.~\ref{fig:ct_sch}(a) corresponds to spindle-like oscillations in the sigma band at $\sim$$14$~Hz \citep{Robinson02}. The peak frequency observed results from resonance of a dominant \textit{SRS} or \textit{SBS} pathway and the boundary is referred to as the sigma instability boundary.
%
%
(iv)~The front boundary corresponds to slow wave oscillations ($<$1 Hz) associated with impending loss of the fixed point stability via a saddle-node bifurcation~\citep{Robinson97,Robinson02,Breakspear06}.

\section{General CTBG system}
\label{sec:asymm}
In this section the results of Sec.~\ref{sec:symm} are generalized by using full asymmetric parameter estimates drawn from the literature and listed in Table~\ref{tab:asymm_params}.
The effect of dopamine depletion on the CTBG model gains is introduced and key loop resonances are investigated as drivers of the peak frequencies present in population activity.
We also investigate whether characteristic activity observed in PD is present in the simplified CTBG system and explore the role of the BG in modulating absence seizure oscillations.

\begin{figure*}[t]
\centering
\includegraphics[width=0.88\textwidth]{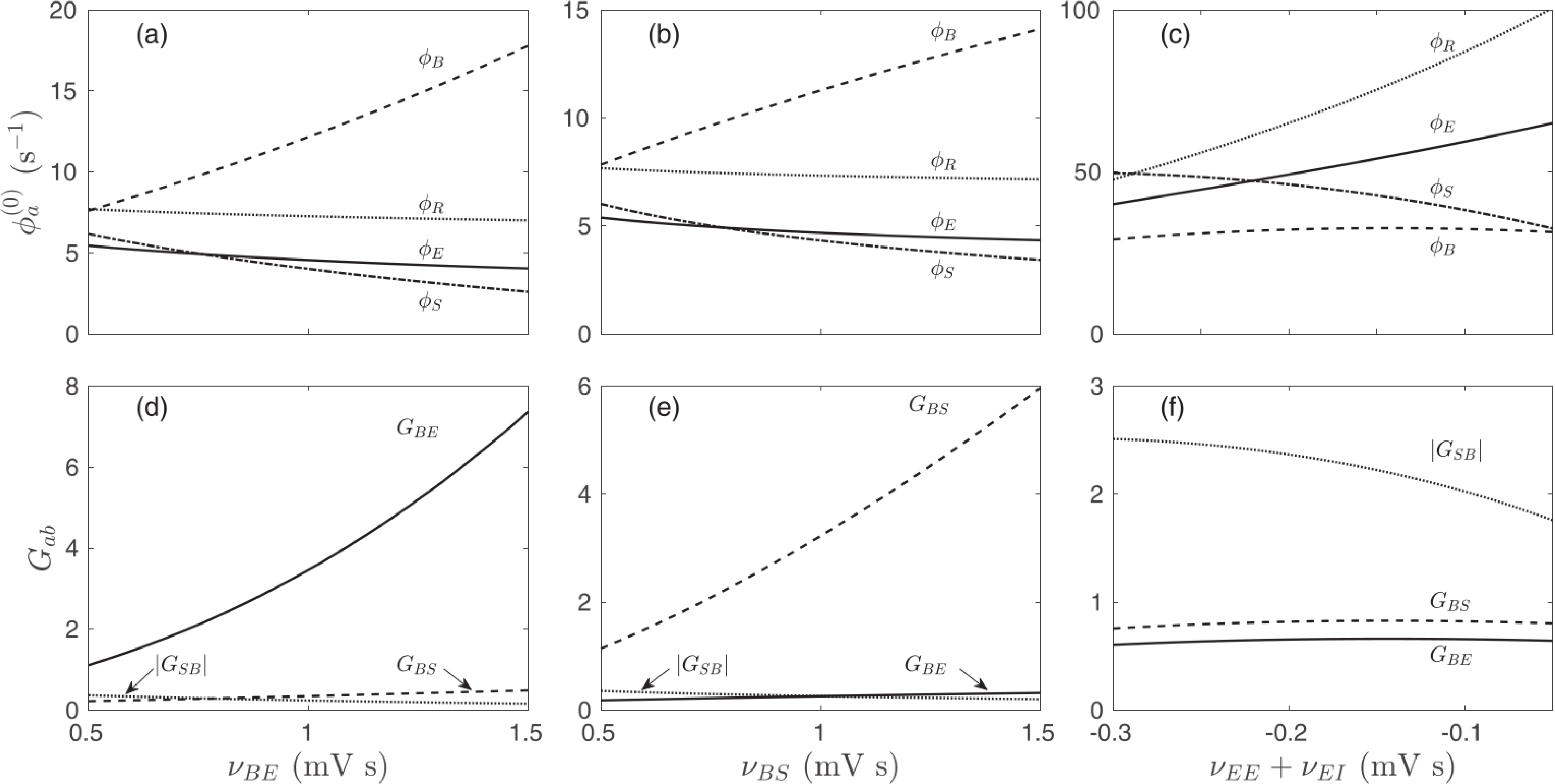}
\caption{\label{fig:ss_pd_firingrates} Dependence of the steady-state population firing rates and BG connection gains on BG afferent coupling and the balance of cortico-cortical excitation and inhibition
coupling. All other connections are set to nominal values defined in Table~\ref{tab:asymm_params} (a) Steady-state population firing rate dependence on cortico-basal ganglia coupling strength. (b)
Steady-state population firing rate dependence on SRN-basal ganglia coupling strength. (c) Steady-state population firing rate dependence on the balance of cortico-cortical
excitation inhibition coupling strengths. (d) BG gain dependence on cortico-basal ganglia coupling strength. (e) BG gain dependence on SRN-basal ganglia coupling strength.
(f) BG gain dependence on the balance of cortico-cortical excitation inhibition coupling strengths.}
\end{figure*}

\subsection{BG parameter estimates}
Here parameters of the full asymmetric BG population, given in Table~\ref{tab:asymm_params}, are determined.
Mean spike activity of GPi nuclei has been measured to be $\sim$$90$~s$^{-1}$ in human patients with PD~\citep{Dostrovsky00,Tang05} and firing rates as high as $\sim$$250$~s$^{-1}$ have been observed~\citep{Nishibayashi11}. 
Because the GPi nuclei serve as a major relay to the thalamus, a maximum firing rate for the simplified BG population is approximated as equivalent to that of the GPi nuclei; i.e., $Q_B^{\rm max} = 200$ s$^{-1}$.
%
%
Van Albada and Robinson (2009) noted that for $Q \ll Q_{\rm max}$, a high maximum firing rate and a low threshold $\theta$ have similar effects on the relatively low steady-state firing rates because Eq.~(\ref{eq:sig}) can be approximated as
\begin{equation}
\label{eq:sig_approx}
\frac{Q^\text{max}}{1+{\rm exp}{[-(V-\theta)/\sigma^\prime]}} \approx Q^\text{max}{\rm e}^{-\theta/\sigma^{\prime}}{\rm e}^{V/\sigma^{\prime}},
\end{equation}
which means a small change $\delta \theta$ to $\theta$ is equivalent to replacing $Q^\text{max}$ by $Q^{\rm max}{\rm e}^{-\delta \theta/\sigma^\prime}$ . Hence, the steady-state dynamics of the CTBG model are largely unchanged by suitably correlated changes in the maximum firing rate and mean threshold potential.

The threshold potential, $\theta_B = 14$~mV, is approximated by numerical exploration of the parameter space to ensure that realistic firing rates, $Q_a \ll Q^\text{max}$, are produced across all populations; i.e., $Q_a \approx 10$ $\text{s}^{-1}$.

A recent study measured the response latency of the globus pallidus internal (GPi) and external (GPe) segments following stimulation of the primary motor cortex (MI) in human adults~\citep{Nishibayashi11}. Their results showed cortically evoked excitation of GPi neurons with a mean response latency of $23\pm 9$~ms. Inhibition was observed after $34\pm 10$~ms and a final excitation after $56\pm 14$~ms.
The delays measured in this study comprise both axonal propagation delays and synaptic and dendritic integration times. In the CTBG model, activity of the simplified BG population represents activity of the GPi segment.
As such, the interpretation of the parameters describing afferent axonal delay and synaptodendritic delay is more general because they incorporate delays of nuclei within the BG that project to the GPi via the direct and indirect pathways.
The dominant component of the axonal delay afferent to the simplified BG results from the characteristic length of the cortico-BG connection which is large because the GPi has a significant spatial separation from the projecting cortical areas; i.e., the motor cortex~\citep{Kandel91}. The characteristic length of the $BE$ connection is assumed to be of the same order of magnitude as that of the corticothalamic connection.
In the absence of specific data, we partition the $23$~ms delay from ~\cite{Nishibayashi11} equally between the axonal delay parameter $\tau_{BE} = 10$~ms and the synaptodendritic rate parameters $\alpha_B = 90$~s$^{-1}$ and $\beta_B = 360$~s$^{-1}$.

\subsection{Approximating dopamine depletion}
\label{subsec:dopa}
As mentioned in the introduction, dopamine depletion is assumed to strengthen the indirect pathway through the BG.
Activity of the simplified BG is considered as an approximation to activity of the GPi/SNr population. 
Since the indirect pathway is also comprised of the D2 striatal, GPe, and STN populations, the effect of these are consolidated into the input gains of the simplified BG population as follows:
\begin{align}
\label{eq:g_be}
G_{BE} = G_{p_2\zeta}G_{\zeta p_1}G_{p_1d_2}G_{d_2e},\\
\label{eq:g_bs}
G_{BS} = G_{p_2\zeta}G_{\zeta p_1}G_{p_1d_2}G_{d_2s},
\end{align}
where the population subscripts are defined in Fig.~\ref{fig:Schem}.
Strengthening the indirect pathway is approximated by an increase in $G_{BE}$.
However, we assume that this increase is not solely attributed to $G_{d_2e}$ in the full system and that the product $G_{p_2\zeta}G_{\zeta p_1}G_{p_1d_2}$ also increases.
Equations~(\ref{eq:g_be}) and (\ref{eq:g_bs}) share this common factor of gains and therefore dopamine depletion in the simplified model results in an increase in $G_{BE}$ but also a relatively smaller increase in $G_{BS}$.
In general, the gains $G_{ab}$ cannot be changed independently since they are functions of steady-state firing rates which are sensitive to changes in the connection parameters.
Figure~\ref{fig:ss_pd_firingrates}(d) shows that increasing the cortico-BG coupling strength $\nu_{BE}$ results in increased values for both the cortico-BG gain $G_{BE}$ and the SRN-BG gain $G_{BS}$.

\subsection{Steady states}
The effect of dopamine depletion on the steady-state population firing rates is determined next.
As mentioned in Sec.~\ref{subsec:dopa}, dopamine loss is approximated as a strengthening of the indirect pathway by increasing the gains $G_{BE}$ and $G_{BS}$. Figures~\ref{fig:ss_pd_firingrates}(a) and (b) show that increasing either the cortico-BG coupling, $\nu_{BE}$, or the SRN-BG coupling, $\nu_{BS}$, tends to increase steady-state firing in the BG population and decrease firing in all other populations.
%
%
In the CTBG model by~\cite{Albada09a}, dopamine depletion increases the steady-state firing rate of the D2 type striatal, GPi and STN populations and decreases all other populations' steady-state firing rates. The D2 type striatal, GPe, STN, and GPi populations constitute the indirect pathway.
Since activity of GPi/SNr population is approximated here as activity of the simplified BG population, the effect of dopamine loss on the steady-state firing rates is consistent between the~\cite{Albada09a} model and the present simplified model.
Furthermore, the result agrees with experiments linking PD effects with hypoactivity of the cortex~\citep{Jenkins92,Monchi04,Monchi07,Martinu12} and the thalamus~\citep{Molnar05}.
%

The mesocortical pathway mediates direct dopaminergic projections of the VTA to the prefrontal cortex~(PFC)~\citep{Gaspar92,Williams93}.
%
\cite{Gulledge01} showed dopamine to have a predominantly inhibitory influence on the PFC due to enhanced synaptic inputs from GABAergic interneurons.
%
Furthermore, both excitatory pyramidal and inhibitory interneurons show increased sensitivity to excitatory inputs in the presence of dopamine~\citep{Gao03,Thurley08}.
Low levels of dopamine in the mesocortical pathway are approximated in the CTBG system by decreasing the intracortical synaptic strengths $\nu_{EE}$, $\nu_{IE}$, and especially $|\nu_{EI}|$ and  $|\nu_{II}|$. 
This agrees with the reduced intracortical inhibition observed in PD patients off medication receiving transcranial magnetic stimulation~\citep{Ridding95}.

In Fig.~\ref{fig:ss_pd_firingrates}(c) the effect of varying the balance between intracortical excitation and inhibition on the steady-state population firing rates is explored.
The connection strengths, $\nu_{EE}$ and $\nu_{EI}$ are initialized at the values given in Table~\ref{tab:asymm_params}. Both are then linearly reduced with $\nu_{EE}$ and $\nu_{IE}$ decreasing by $\Delta\nu$, and $|\nu_{EI}|$ and $|\nu_{II}|$ by $2\Delta\nu$ at each step.
This reduction of intracortical inhibition relative to excitation tends to increase $Q_E$ and $Q_{R}$, and decrease $Q_{S}$.
The BG population's steady-state firing rate is only weakly affected by changes in the intracortical excitation-inhibition balance over the parameter ranges considered.
\begin{table}[h!]
\caption{\label{tab:asymm_params} Nominal parameters for the asymmetric system, adapted
from \cite{Abeysuriya14}.}
\begin{tabularx}{\columnwidth}{ l X X r }
\hline 
Quantity & & Value & Unit\\ 
\hline
$r$ 						&	& $80$ 		& mm			\\
$\sigma^\prime$					&	& $3.3$		& mV			\\
$\phi_{N}^{(0)}$				&	& $1$ 		& $\text{s}^{-1}$	\\
$\gamma_{E}$					&	& $116$ 	& $\text{s}^{-1}$	\\
$\alpha_{E,R,S}$				&	& $45$ 		& $\text{s}^{-1}$	\\
$\beta_{E,R,S}$					&	& $180$ 	& $\text{s}^{-1}$	\\
$\alpha_B$					&	& $90$ 		& $\text{s}^{-1}$	\\
$\beta_B$					&	& $360$	 	& $\text{s}^{-1}$	\\
$\tau_{RE},\tau_{SE}$				&	& $45$	 	& ms\\
$\tau_{ES}$					&	& $35$	 	& ms\\
$\tau_{BE}$					&	& $10$	 	& ms\\
$Q_E^{\rm max}$, $Q_S^{\rm max}$,$Q_R^{\rm max}$&	& $300$ 	& $\text{s}^{-1}$	\\
$Q_B^{\rm max}$					&	& $200$ 	& $\text{s}^{-1}$	\\
$\theta_E$, $\theta_B$				&	& $14$ 		& mV			\\
$\theta_R$, $\theta_S$				&	& $13$ 		& mV			\\
$\nu_{EE}$					&	& $1.6$ 	& mV s			\\
$\nu_{EI}$ 					&	& $-1.9$ 	& mV s			\\
$\nu_{ES}$ 					&	& $0.4$ 	& mV s			\\
$\nu_{RE}$ 					&	& $0.15$ 	& mV s			\\
$\nu_{RS}$	 				&	& $0.03$ 	& mV s			\\
$\nu_{BE}$ 					&	& $0.08$ 	& mV s			\\
$\nu_{BS}$ 					&	& $0.1$ 	& mV s			\\
$\nu_{SE}$ 					&	& $0.8$ 	& mV s			\\
$\nu_{SR}$					&	& $-0.4$ 	& mV s			\\
$\nu_{SB}$ 					&	& $-0.2$ 	& mV s			\\
$\nu_{SN}$ 					&	& $0.5$ 	& mV s			\\
\hline
\end{tabularx}
\end{table}
%

\begin{figure*}[t]
\centering
\includegraphics[width=0.88\textwidth]{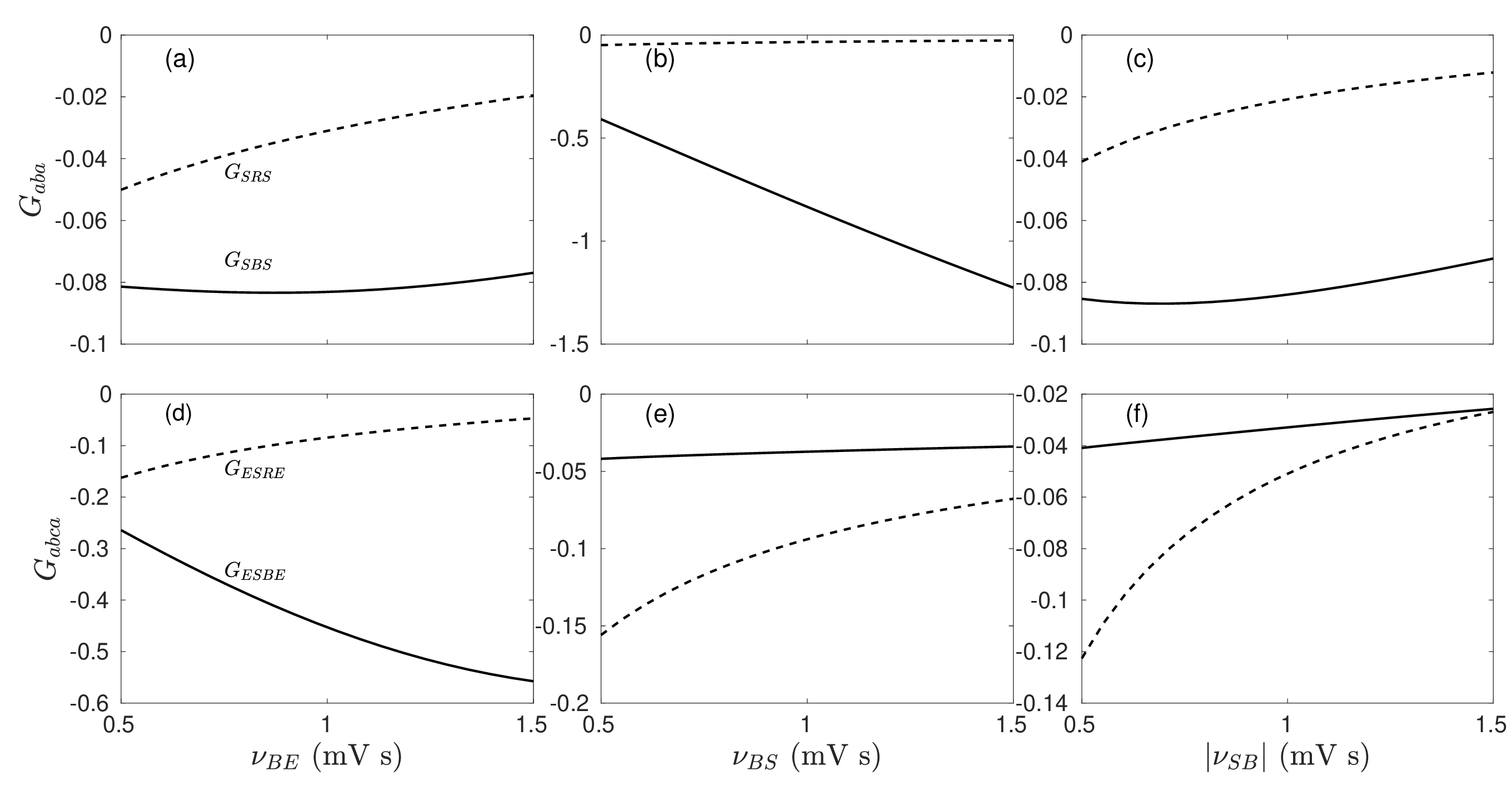}
\caption{\label{fig:loop_gains} Dependence of loop gains involving the BG or TRN on the BG connection strengths. All other connections are set to nominal values defined in Table~\ref{tab:asymm_params}. Intrathalamic $G_{SBS}$~(solid) and BG-thalamic $G_{SRS}$~(dashed) dependence on (a) cortico-BG coupling strength, (b) SRN-BG coupling strength, and (c) BG-SRN coupling strength. Dependence of $G_{ESBE}$~(solid) and $G_{ESRE}$~(dashed) on (d) cortico-BG coupling strength, (e) SRN-BG coupling strength, and (f) BG-SRN coupling strength.}
\end{figure*}

\subsection{Alpha resonance}
\label{subsec:alpha}
The alpha rhythm~($\sim$$10$~Hz) is a key feature of the CT model. Strong alpha oscillations can be produced in the model that are comparable to those observed clinically in tonic-clonic~(grand mal) seizure spectra~\citep{Robinson01,Robinson02,Breakspear06}.
As mentioned in Sec.~\ref{sec:symm}, the alpha rhythm results from strong cortico-thalamo-cortical coupling, $G_{ESE}$. Cortical activity has a net excitatory effect on the SRN population which then projects excitation back to the cortex after a time
\begin{equation}
T \approx \tau_{SE}+\tau_{ES} + \frac{2}{\gamma_{E}}  + \frac{2}{\alpha_{E}} + \frac{2}{\beta_{E}}\\.
\end{equation}
As the system approaches the alpha instability boundary, described in Sec.~\ref{sec:symm}, the
alpha resonance becomes less damped and thus stronger.
%
%
Inclusion of the BG in this model does not affect the temporal properties of the \textit{ESE} loop, which is responsible for the alpha resonance, and thus does not significantly affect the peak frequency observed.

%
%
Approximating dopamine depletion by increasing the cortico-basal ganglia gain $G_{BE}$ will result in an increased $|G_{ESBE}|$ loop gain, as shown in Fig.~\ref{fig:loop_gains}(d). This increases $|G_{ESRE}+G_{ESBE}|$ and tends to move the system away from the alpha instability boundary. The effect can be interpreted as a reduced likelihood of tonic-clonic (grand mal) seizures and agrees with experimental results on generalized tonic-clonic seizures in the rat brain~\citep{Velivskova96,Dybdal00} where the indirect pathway demonstrated anti-epileptic effects.

\subsection{Generalized delta/theta resonance}
\label{subsec:theta}
Another key feature of the CT model is the presence of global activity at $2$--$5$~Hz that can describe EEG spectra recorded during absence~(petit mal) seizure~\citep{Robinson01,Robinson02,Breakspear06}.
The oscillations in the model result from a resonance in the cortico-thalamo-cortical loop, referred to as the delta/theta resonance, where cortical activity projects onto the SRN via the TRN, and then feeds back into the cortex (\textit{ESRE} loop). Similarly to the alpha resonance, proximity to the delta/theta instability boundary reduces damping of the delta/theta resonance.
The \textit{ESRE} loop provides negative feedback to the cortex as TRN activity inhibits the SRN~($\nu_{SR} < 0$). A positive net gain is achieved on the second pass and thus a frequency peak is observed corresponding to the total time required for two circuits of the \textit{ESRE} loop~\citep{Robinson02}.
%

\begin{figure}[h!]
\includegraphics[width=0.47\textwidth]{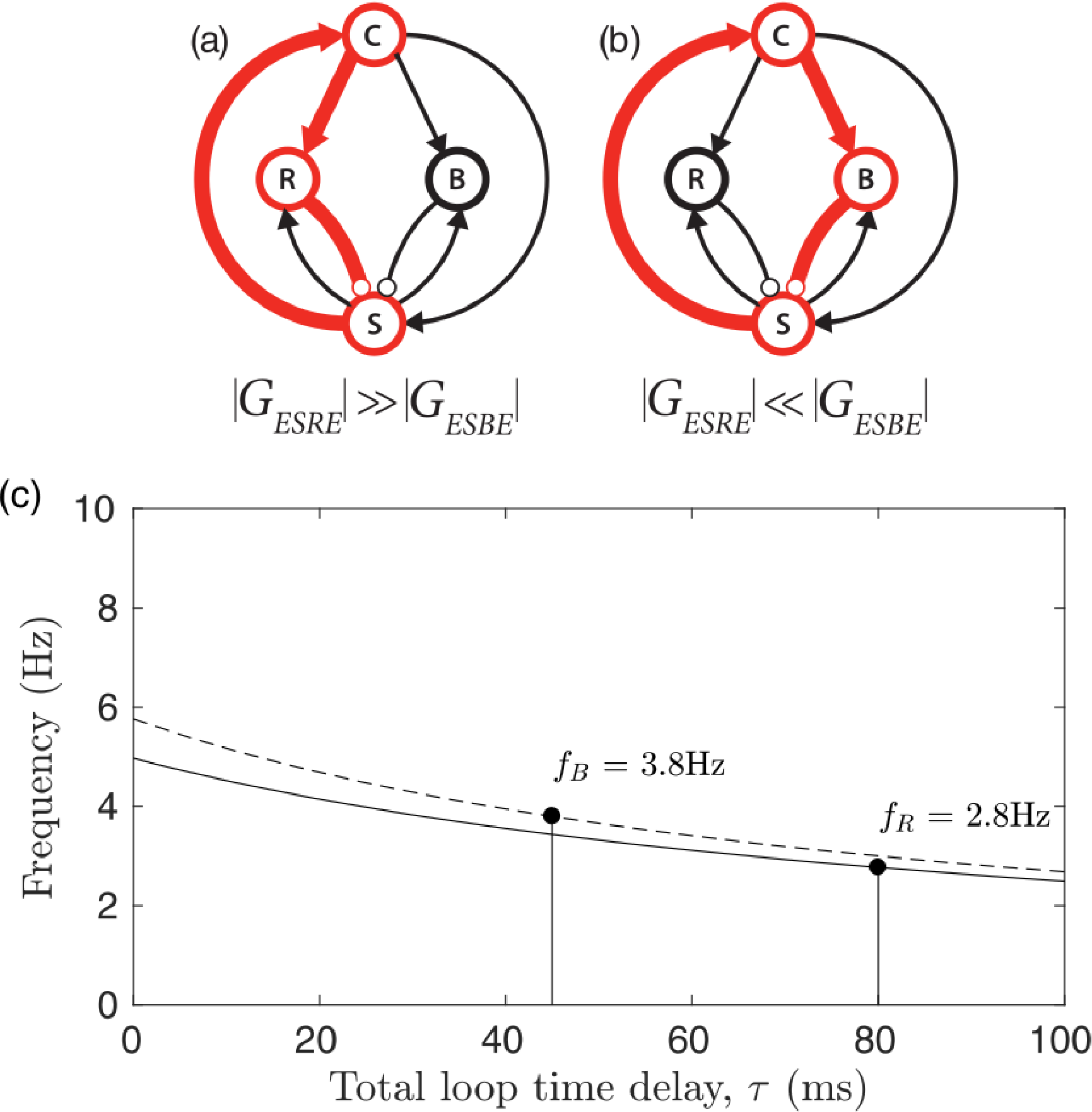}
\caption{\label{fig:theta} Properties of the delta/theta resonance. (a) Schematic of a dominant \textit{ESRE} loop near the delta/theta instability boundary. (b) Schematic of a dominant \textit{ESBE} loop near the delta/theta instability boundary. (c) Predicted delta/theta resonance frequencies for the two pure cases, (a) and (b), are plotted as functions of the net delay in the loop, $\tau$; via the basal ganglia $\tau = \tau_{ESBE}$~(dash) and via the reticular nucleus $\tau = \tau_{ESRE}$ (solid). Frequencies prescribed by the period of the two isolated pathways when using the time delays from Table~\ref{tab:asymm_params} are also shown (heavy dots).}
\end{figure}

The \textit{ESBE} loop, analogously to the \textit{ESRE} loop, provides negative feedback to the cortex via the BG and SRN populations.
Each of these parallel loops supports an oscillation with a period determined by the combination of axonal and synaptodendritic response delays.
The overall delta/theta resonance frequency of the system then results from the combined effects of activity traversing the two parallel pathways. In order to estimate this peak frequency each loop is first considered separately. For $G_{ESBE} = 0$,
\begin{align}
\label{eq:theta_thal}
T_{\theta} \approx
2\left(\tau_{RE} + \tau_{SR} + \tau_{ES}\right)  + \frac{4}{\gamma_{E}} + 6\left[\frac{1}{\alpha_{E}} + \frac{1}{\beta_{E}}\right].
\end{align}
while for $G_{ESRE} = 0$,
\begin{align}
\label{eq:theta_bg}
T_{\theta} &\approx 2\left(\tau_{BE} + \tau_{SB} + \tau_{ES}\right) + \frac{4}{\gamma_{E}} 
+ 4\left[\frac{1}{\alpha_{E}} + \frac{1}{\beta_{E}}\right] \nonumber\\
& \qquad + 2\left[\frac{1}{\alpha_{B}} + \frac{1}{\beta_{B}}\right],
\end{align}
Equation~(\ref{eq:theta_thal}) reproduces the previous result of the CT model~\citep{Robinson02} where the \textit{ESRE} loop has a resonant frequency near $3$~Hz.

Figure~\ref{fig:theta}(c) shows the predicted delta/theta frequencies of the two pure \textit{ESRE} and \textit{ESBE} cases, given by Eq.~(\ref{eq:theta_bg}) and (\ref{eq:theta_thal}) respectively. The frequencies depend on the net axonal delay, $\tau$, where $\tau$ is either a corticothalamic axonal delay $\tau_{ESRE}$ or a cortico-basal ganglia-thalamic axonal delay $\tau_{ESBE}$. 
In our model, the axonal and synaptodendritic delays for the \textit{ESBE} pathway are smaller than for the \textit{ESRE} pathway. Thus, the period of the \textit{ESBE} loop is lower than that of the \textit{ESRE} loop. For the combined system, the frequency is between the values supplied by Eq.(\ref{eq:theta_bg}) and (\ref{eq:theta_thal}).
%

As mentioned in Sec.~\ref{subsec:dopa}, in approximating dopamine depletion we increase $G_{BE}$. To achieve this we increase the connection strength $\nu_{BE}$. Figure~\ref{fig:loop_gains}(d) shows that, using nominal parameters from Table~\ref{tab:asymm_params}, increased $\nu_{BE}$ causes an increase in $|G_{ESBE}|$ and, to a lesser extent, a corresponding decrease in $|G_{ESRE}|$.
This tends to move the CTBG system towards the delta/theta instability boundary and results in weakened damping of the \textit{ESBE} loop resonance.
Enhanced $4$~Hz oscillations are then present in cortical, BG, and SRN population firing rates, which are similar to the frequencies typical of tremulous PD~\citep{Brown01,Wang05}.
%
The oscillations are also consistent with the findings of the more detailed CTBG model by~\cite{Albada09b} who demonstrated activity about $5$~Hz was sustained by a dominant indirect BG pathway.
The $4$~Hz activity is comprised of coordinated action between cortical and SRN populations, and BG nuclei that form the indirect pathway. This result can thus plausibly explain the correlated LFP fluctuations about these frequencies observed between thalamic nuclei and the BG~\citep{Brown01,Wang05,Smirnov08,Tass10}.
It also provides an explanation for the observed coherence between LFP fluctuations in the BG and EEG measured over the motor cortex~\citep{Marsden01,Williams02}.

Absence seizure like~$\sim$$3$~Hz activity in the model results from a dominant $G_{ESRE}$ loop gain~\citep{Robinson01,Robinson02,Breakspear06}.
The increased proximity to the delta/theta instability boundary in dopamine depleted states suggests that a smaller change in the strength of $G_{ESRE}$, relative to non-parkinsonian states, is required to drive this enhanced 3~Hz activity.
This agrees with several studies reporting an increased likelihood of absence seizures after dopamine depletion~\citep{Danober98,Deransart98,Deransart00,De00,Mid01} and decreased dopamine metabolism in human generalized epilepsy~\citep{Bouilleret05}.

\subsection{Generalized sigma resonance}
\label{subsec:sigma}
\begin{figure}[h!]
\includegraphics[width=0.50\textwidth]{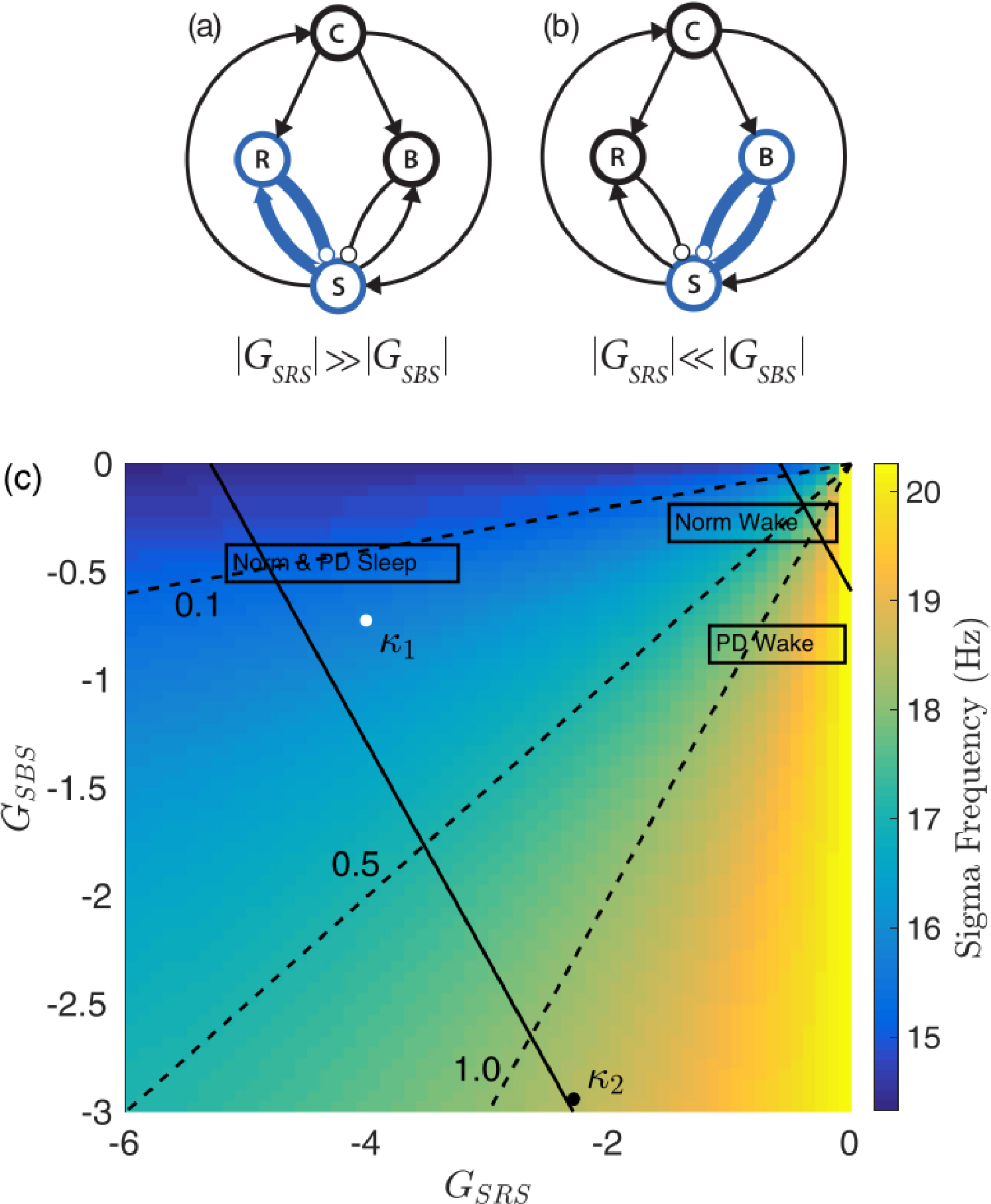}
\caption{\label{fig:sigma}
Properties of the sigma resonance. (a) Schematic of a dominant \textit{SRS} loop
near the sigma instability boundary. (b) Schematic of a dominant \textit{SBS} loop near
the sigma instability boundary. (c) Dependence of the sigma resonant frequency on the loop gains $G_{SRS}$ and $G_{SBS}$. To allow for repartitioning in the present model, the sleep and wake estimates for the CT system from \cite{Abeysuriya14} are plotted as the lines $G_{SRS}+G_{SBS} = -5.3$ for CT sleep spindles and $G_{SRS}+G_{SBS} = -0.6$ for CT wake. The dotted lines show values of the ratio of $G_{SBS}$ compared to $G_{SRS}$. The parameters used in Fig.~\ref{fig:spectra}(d) and (e) are marked as $\kappa_1$~(white dot) and $\kappa_2$~(black dot) respectively. The rectangles show the approximate values of the gains during sleep stage 2 and wake for both normal and PD conditions.}
\end{figure}

Sleep spindles are oscillations at 12--15~Hz characteristic of EEG recordings made during sleep stage 2 and anesthesia~\citep{Wolter06,Ferenets06}.
An important feature of the previous CT model~\citep{Robinson02} is its ability to generate spindle frequencies within the intrathalamic \textit{SRS} loop, see Fig.~\ref{fig:sigma}(a).
The CTBG system contains an additional basal ganglia-thalamic loop, \textit{SBS}, that is analogous to the \textit{SRS} loop, as shown in Fig.~\ref{fig:sigma}(b).
The peak frequencies expected in proximity to the more general sigma instability boundary are analyzed next.

%
%
\cite{Robinson02} explored a parameter regime of dominant intrathalamic dynamics in the CT model. Roots of the dispersion relation exist where $1 - L_{S}L_{R}G_{SRS} = 0$. At the boundary of the stability zone, solutions to this relation are purely real and thus satisfied by $\omega = \omega_\sigma = \sqrt{\alpha\beta}$; i.e., a pure spindle instability.
%

For the CTBG model, a more general case is considered in which intrathalamic (\textit{SRS}) and basal ganglia-thalamic (\textit{SBS}) loop dynamics dominate. In this case, $1 - J_{SRS} - J_{SBS}$ is the dominant term in Eq.~(\ref{eq:disp}) with roots depending on the relative strengths of the gains,
$G_{SRS}$ and $G_{SBS}$, as well as the time delays associated with each loop.
Expanding this term using Eq. (\ref{eq:Ja1a2}), roots of (\ref{eq:disp}) satisfy
%
\begin{equation}
\label{eq:Ls}
1 - L_{S}\left[L_{R}G_{SRS}e^{i\omega\tau_{SRS}} + L_{B}G_{SBS}e^{i\omega\tau_{SBS}}\right] = 0.\\
\end{equation}
%

\begin{figure*}[t!]
\centering
\includegraphics[width=0.5\textwidth]{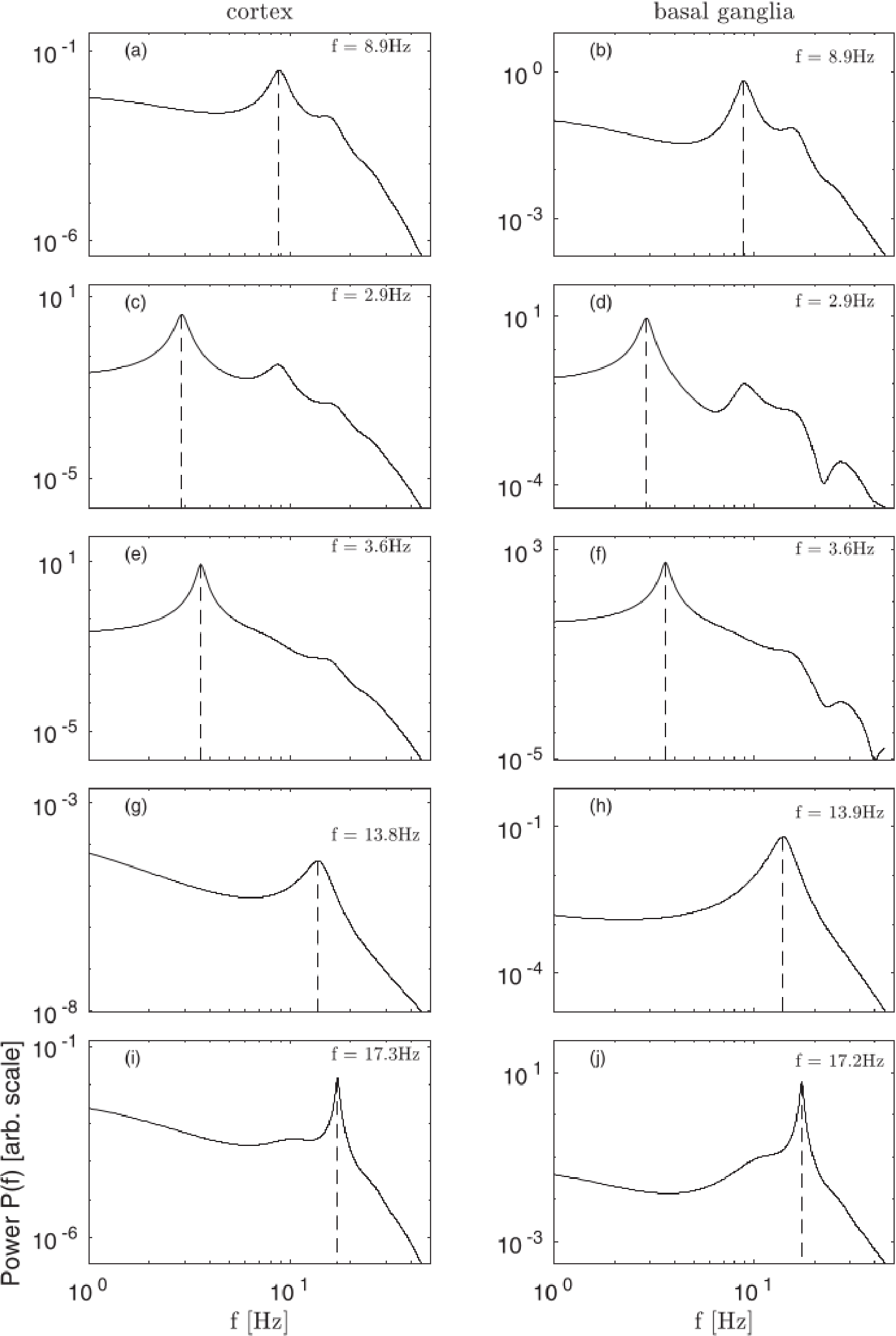}
\caption{\label{fig:spectra}
Analytical spectra of cortical (1st column) and BG (2nd column) population
activity calculated from Eq.~(26). The peak frequency predictions are indicated in
each plot. Each row corresponds to a different set of connection strengths. (a) and
(b) show the alpha resonance for $G_{ESE} = 3.15$, $G_{ESRE}=-0.15$, and $G_{ESBE}=-0.14$. (c) and (d) show the delta/theta resonance for $G_{ESE} = 2.55$, $G_{ESRE}=-4.61$, and $G_{ESBE}=-0.88$. (e) and (f) show the delta/theta resonance with $G_{ESE} = 2.20$, $G_{ESRE}=-2.18$, and $G_{ESBE}=-3.44$. (g) and (h) show the sigma resonance with $G_{SRS}=-4.01$, and $G_{SBS}=-0.73$. (i) and (j) show the sigma resonance with $G_{SRS}=-2.29$, and $G_{SBS}=-2.94$.}
\end{figure*}

\begin{figure*}[t!]
\centering
\includegraphics[width=0.8\textwidth]{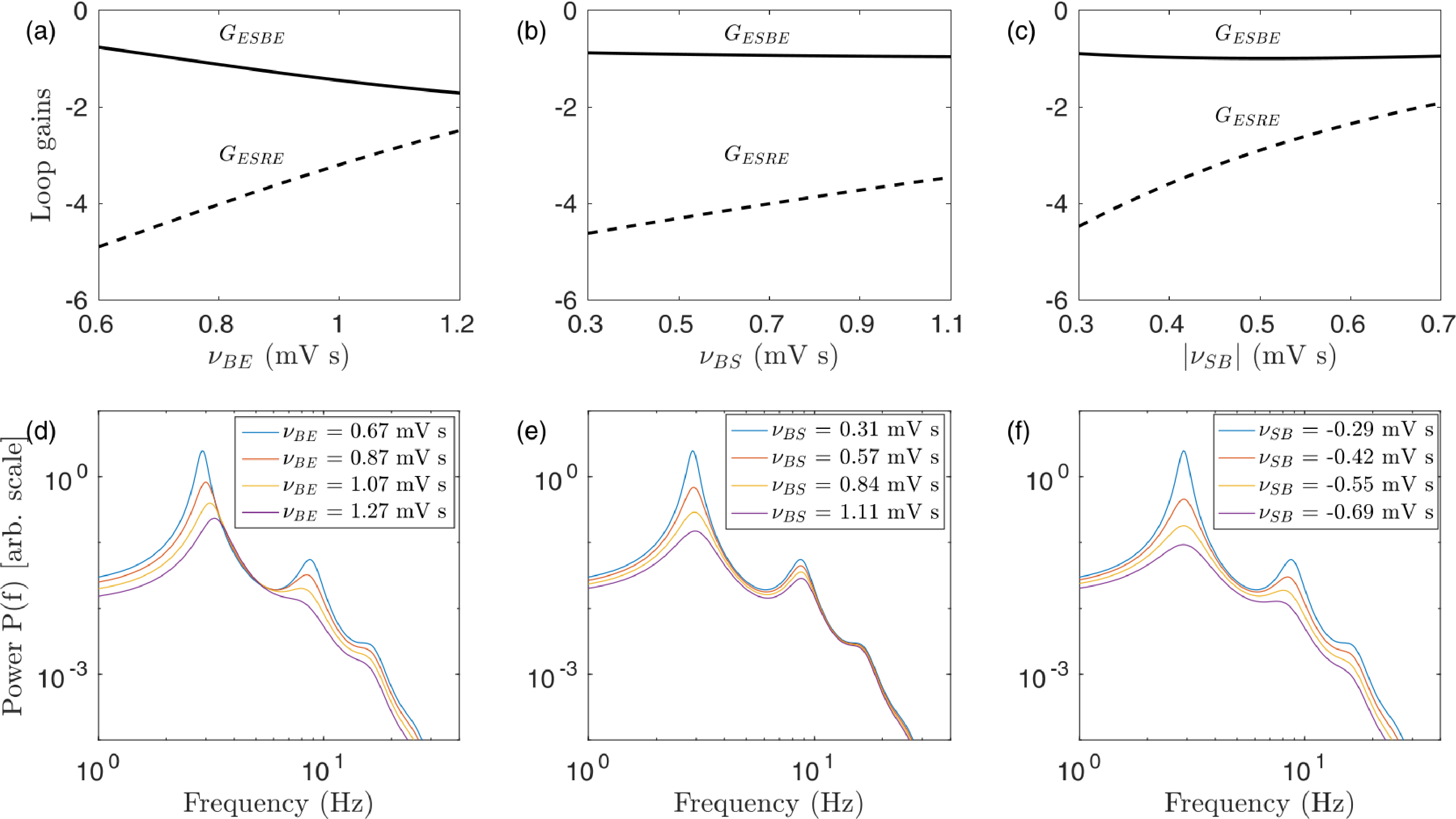}
\caption{\label{fig:theta_supp} Suppression of $\sim$$3$~Hz activity by strengthening the afferent and efferent BG connections. All other connections are set to values defining a dominant $G_{ESRE}$ loop gain. (a) BG and TRN corticothalamic loop gain dependence on cortico-BG coupling strength. (b) BG and TRN corticothalamic loop gain dependence on SRN-BG coupling strength. (c) BG and TRN corticothalamic loop gain dependence on BG-SRN coupling strength. (d) Power spectrum for increasing values of cortico-BG coupling strength. (e) Power spectrum for increasing values of SRN-BG coupling strength. (f) Power spectrum for increasing values of BG-SRN coupling strength.}
\end{figure*}

%
%
Using a numerical bisection method, solutions to Eq.~(\ref{eq:Ls}) are
found at the instability boundary where roots $\omega$ are purely real
and the results are presented in Fig.~\ref{fig:sigma}(c).
%
The upper left region of~\ref{fig:sigma}(c) defines states of a dominant $SRS$ loop represented by  the schematic in Fig.~\ref{fig:sigma}(a). The lower right region defines states of a dominant $SBS$ loop represented by the schematic in Fig.~\ref{fig:sigma}(b). Figures~\ref{fig:spectra}(d) and (e) show examples of activity produced by dominant $G_{SRS}$ and $G_{SBS}$ sigma resonances, respectively.
As mentioned in Sec.~\ref{subsec:dopa}, the connection gains are interdependent so $G_{SRS}$ and $G_{SBS}$ can not be varied independently in general. The effects of the BG connection strengths, $\nu_{BE}$, $\nu_{BS}$, and $\nu_{SB}$ on $G_{SRS}$ and $G_{SBS}$ are shown in Figs~\ref{fig:loop_gains}(a)-(c).
The connection gains previously used to describe the TRN population in the CT model are repartitioned between the TRN and BG in the present model, as discussed in Sec.~\ref{subsec:repart}.
Thus, sleep and wake CT parameters estimates from \cite{Abeysuriya14} are plotted in Fig.~\ref{fig:sigma}(c) as straight lines representing constant sums $G_{SRS}+G_{SBS}$; i.e., the CT sleep line $G_{SRS}+G_{SBS} = -5.3$ and the CT wake line $G_{SRS}+G_{SBS} = -0.6$.

%
%
Parkinsonian patients have sleep spindle frequencies around $12$--$15$~Hz that are similar to those observed in healthy controls~\citep{Christensen15,Latreille15}.
Sleep states for both healthy and PD patients must thus lie in the upper left part of Fig.~\ref{fig:sigma}(c) where the ratio $G_{SBS}/G_{SRS} \approx 0.1$. Thus, the spindle frequency is primarily driven by a weakly damped intrathalamic loop resonance $SRS$, which is consistent with results of the CT model~\citep{Robinson02,Abeysuriya14}.
Furthermore, the change in $|G_{SRS}|$ between sleep stage 2 and wake, and vice versa, likely remains very similar for PD and healthy states since both require a dominant $G_{SRS}$ gain to generate spindles.

%
%
Wake values for $|G_{SBS}|$ and $|G_{SRS}|$ present the most significant difference between PD and healthy controls.
For spindle generation, $|G_{SRS}+G_{SBS}|$ is near the sleep estimate from \cite{Abeysuriya14} where the sigma resonance is dominant and results in a pronounced peak in the spectra of population activity.
The $\sim$$20$~Hz activity in PD is not a sharp peak in the activity spectra. Instead these oscillations describe a frequency of peak coherence between the BG and thalamic population activity and which becomes strengthened in PD.
Thus, values for the total gain $|G_{SBS} + G_{SRS}|$ are smaller than those for spindle generation.
As mentioned in Sec.~\ref{subsec:dopa}, in approximating dopamine loss we increase $G_{BS}$. Figure~\ref{fig:loop_gains}(b) shows that increasing $\nu_{BS}$ increases $G_{SBS}$ and will tend to move wake states of the CTBG system towards the bottom right of Fig.~\ref{fig:sigma}(c) where the ratio $G_{SBS}/G_{SRS}$ is close to $1$.
The increased proximity to the sigma boundary will weaken the damping of the \textit{SBS} loop resonance and drive enhanced beta band oscillations at $\sim$$20$~Hz.
As in the case of the delta/theta resonance in PD, the $\sim$$20$~Hz activity results from coordinated action between the SRN and BG and is thus consistent with the presence of synchronized $\sim$$20$~Hz activity observed between the BG and thalamic nuclei~\citep{Brown01,Brown05,Levy02,Timmermann03,Kuhn05}.
%

\begin{figure*}[t!]
\centering
\includegraphics[width=0.8\textwidth]{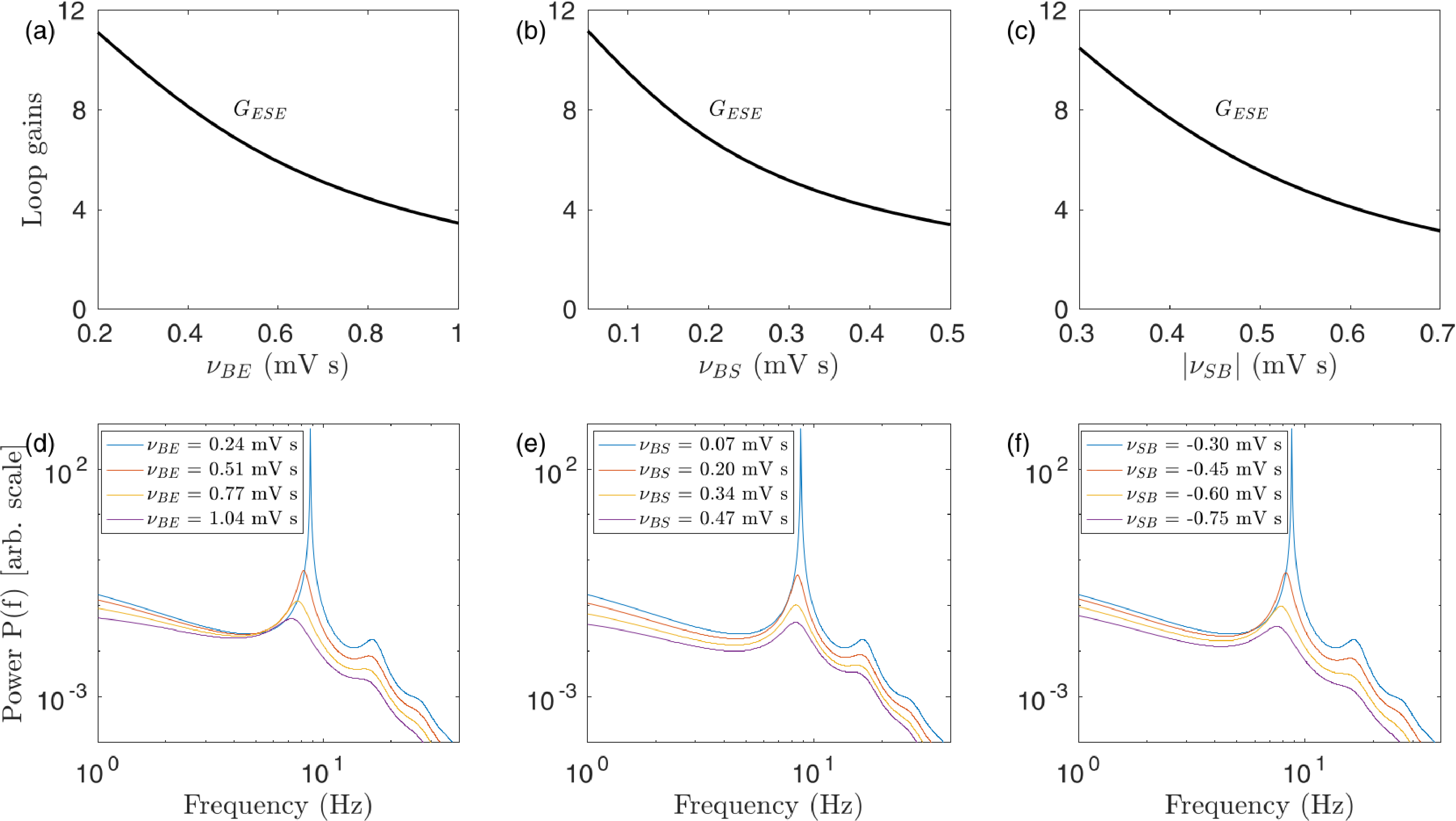}
\caption{\label{fig:alpha_supp} Suppression of $\sim$$10$~Hz activity by strengthening the afferent and efferent BG connections. All other connections are set to values defining a dominant $G_{ESE}$ loop gain.
(a) BG and TRN corticothalamic loop gain dependence on cortico-BG coupling strength. (b) BG and TRN corticothalamic loop gain dependence on SRN-BG coupling strength.
(c) BG and TRN corticothalamic loop gain dependence on BG-SRN coupling strength. (d) Power spectrum for increasing values of cortico-BG coupling strength. (e) Power
spectrum for increasing values of SRN-BG coupling strength. (f) Power spectrum for increasing values of BG-SRN coupling strength.}
\end{figure*}

\subsection{Analytical power spectrum}
The frequency response of the excitatory population to low intensity white noise input $\phi_N$ is generated using the transfer function Eq.~(21) and the results are shown in column one of Figure~\ref{fig:spectra}.
The frequency response of the BG population with respect to the
excitatory population is given by
\begin{equation}
\frac{\phi_B(\omega)}{\phi_E(\omega)} = J_{BE} + \frac{J_{BS}}{J_{ES}}\left[(1-J_{EI})\left(1-\frac{i\omega}{\gamma_E} \right)^2 - J_{EE} \right].
\end{equation}
and is used to produce column two of Fig.~7. Each row of Fig.~7 is
produced using a unique set of connection strengths and demonstrates the model's ability to generate the alpha, delta/theta, and beta band activity predicted in Sections 4.4--4.6. The $\sim$3 Hz (shown
in Fig.~7(a) and (b)) and $\sim$10~Hz (shown in Fig. 7(c) and(d)) correspond to states near absence and tonic-clonic seizure. These states
are in close proximity to the instability boundary and represent
precursive activity to the broadband oscillations generated as the
boundary is traversed, as explored in Breakspear et al. (2006).
The trough in Fig. 7(d) at around $\sim$22~Hz is caused by a zero in
Eq.~(40) and occurs at a significantly lower power than the dominant $\sim$3~Hz peak.
%
%
%
%


\subsection{BG modulation of absence seizure}
\label{subsec:absence}
%
%
As discussed in Sec.~\ref{subsec:theta}, the delta/theta resonance, driven by a strong $G_{ESRE}$ loop gain, is capable of generating $\sim$$3$~Hz oscillations.
In this case, the system is near the instability boundary but still linearly stable due to weak damping of the delta/theta resonance. At or beyond the delta/theta instability boundary, the fixed point of the system loses stability and represents transition to absence seizure~\citep{Robinson02,Breakspear06}.
Figure~\ref{fig:spectra}(b) shows example spectra near the delta/theta instability boundary.

Several recent studies have aimed to explore the role of the BG in absence seizure.
\cite{Chen14} and~\cite{Hu15} extended the \cite{Albada09a} CTBG model by introducing an additional connection coupling GPi/SNr activity to the TRN that parallels the existing GPi/SNr-SRN connection. The model was configured to generate absence seizure oscillations and then these oscillations were modulated by changing GPi/SNr population activity.
Both studies demonstrate absence seizure suppression resulting from increased activity of the GPi/SNr by strengthening afferent coupling of STN activity, which is equivalent to increasing $\nu_{BE}$ and $\nu_{BS}$ in the simplified model.
Fig.~8(a)--(c) show the dependence of the gains $G_{ESRE}$ and $G_{ESBE}$
on the afferent and efferent BG connection strengths when the
system is near the delta/theta instability boundary. In Fig.~8(d)--(f), the prominent peak at $\sim$3~Hz is diminished by larger values of each BG connection. The effect is due to a resulting reduction
of $|G_{ESRE}|$, as shown in Fig.~8(a)--(c). In contrast $|G_{ESBE}|$ increases
over the same range of connection strengths but at a much lower
rate than the decrease of $|G_{ESRE}|$.
%
%
%
%
%
This means that $|G_{ESRE} + G_{ESBE}|$ is decreasing. Since \mbox{$G_{ESE} < |G_{ESRE} + G_{ESBE}|$} near the delta/theta instability boundary, the effect is to move the system away from this boundary, resulting in increased damping of the delta/theta resonance and suppression of the $\sim$$3$~Hz oscillations. The peak in
Fig.~8(a) is shifted to the right for larger values of $\nu_{BE}$ because $G_{ESBE}$
is also increasing which represents the $\sim$4~Hz generating loop.

The suppression of the $\sim$3~Hz activity is independent of which
BG connection is strengthened and instead depends on weakening a combined dominance of $|G_{ESRE} + G_{ESBE}|$ relative to $G_{ESE}$.
The simplified model is thus consistent with both \cite{Chen14} and~\cite{Hu15}, and provides new insight into their results.
%



\subsection{BG modulation of tonic-clonic seizure}
Tonic-clonic seizure states of the model occur as the system traverses the alpha instability boundary~\citep{Breakspear06}. A dominant alpha resonance, defined by a large $G_{ESE}$ gain, produces enhanced $\sim$$10$~Hz activity. 
As described in Sec.~\ref{subsec:alpha}, strengthening $G_{BE}$ by increasing $\nu_{BE}$, will tend to move the system away from the alpha instability boundary shown in Fig.~2(a). This is due to a resulting increase in $G_{ESBE}$ but also a decrease in the dominant $G_{ESE}$ gain.
%
%
%
Fig.~\ref{fig:alpha_supp}(a)--(c) show that $G_{ESE}$ can be decreased by a
corresponding increase in any of the BG connection strengths.
%
In Fig.~\ref{fig:alpha_supp}(d)--(f), the prominent $\sim$10~Hz peak driven by a large $G_{ESE}$ gain is shown to diminish for larger values of the BG connection
strengths.
%

%
Several studies have found the BG to play an important role
in tonic-clonic seizure \citep{Blumenfeld09, Ciumas06, Luo11, Nersesyan04}. An increased threshold to flurothyl-induced tonic-clonic seizure has been shown by
high frequency stimulation of the STN \citep{Lado03} and by
infusion of an NMDA antagonist in the SNR \citep{Velisek06}.
These studies are consistent with the suppressive effect of the BG
in the CTBG model and support the reported efficacy of tonic-clonic seizure treatments targeting neural circuits involving the BG.


\subsection{TRN modulation of parkinsonian states}
%
Parallel to the results of generalized epilepsies in Sections 4.8 and 4.9, we are motivated to consider the possible role of the TRN in modulating parkinsonian states.

Sleep states in the CTBG model are described by strong afferent and efferent TRN connections and high TRN activity relative to values in wake~\citep{Robinson02,Abeysuriya14}.
Fig.~10(a)--(c) demonstrate that strengthening $\nu_{RE}$, $\nu_{RS}$ or $|\nu_{SR}|$, decreases a dominant $G_{ESBE}$ loop gain that drives the weakly damped $\sim$4~Hz oscillations. In Fig.~10(d)--(f) the prominent $\sim$4~Hz peak is diminished for larger values of these TRN connection strengths.

%
Circadian fluctuations of motor symptoms often occur in PD, with less severe motor dysfunction in the early morning than in the afternoon~\citep{Marsden82,Fahn82} and parkinsonian tremor and rigidity typically disappear during sleep, with minor tremor occasionally persisting in slow-wave sleep~\citep{Broughton67,Stern68}.
%
%
These observations are consistent with sleep states of the CTBG model having a suppressive effect on pathological activity in PD. The alleviation of motor dysfunction in the early morning may be caused by intermediate values for the TRN connections during the sleep-to-wake transition and moderate TRN activity relative to wake levels.   
%


%
%
%

\begin{figure*}[t!]
\centering
\includegraphics[width=0.8\textwidth]{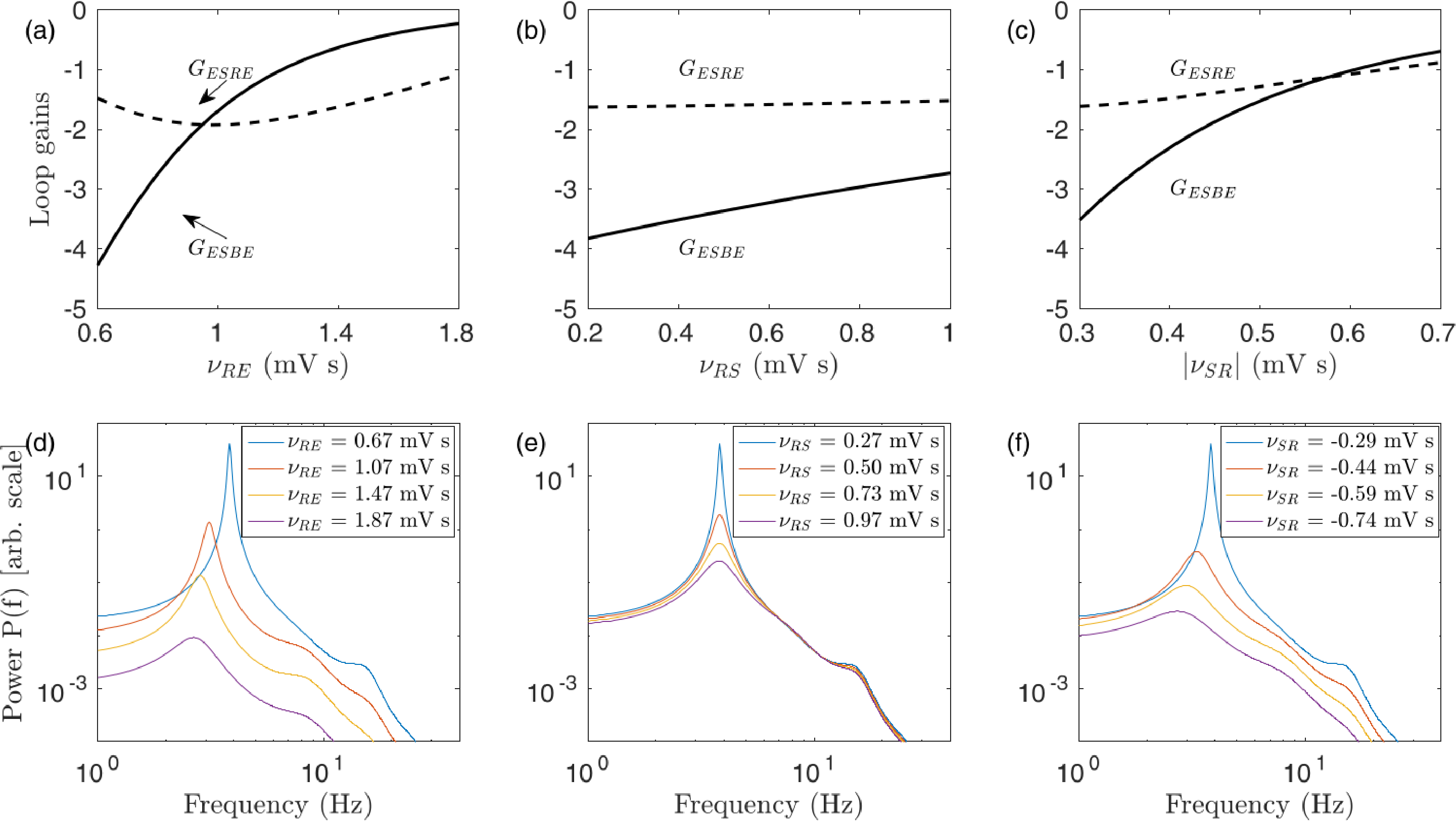}
\caption{\label{fig:4Hz_supp} Suppression of $\sim$4~Hz activity by strengthening the afferent and efferent TRN connections. All other connections are set to values defining a dominant $G_{ESBE}$ loop gain. (a) BG and TRN corticothalamic loop gain dependence on cortico-TRN coupling strength. (b) BG and TRN corticothalamic loop gain dependence on SRN-TRN coupling
strength. (c) BG and TRN corticothalamic loop gain dependence on TRN-SRN coupling strength. (d) Power spectrum for increasing values of cortico-TRN coupling strength.
(e) Power spectrum for increasing values of SRN-TRN coupling strength. (f) Power spectrum for increasing values of TRN-SRN coupling strength.}
\end{figure*}

\section{Discussion}
\label{sec:disc}
We have proposed a physiologically-based neural field model of
the CTBG system, and have used it to explore PD and epilepsy in a
unified framework. The model simplifies the approach of \cite{Albada09a} by approximating the BG as a single effective population whose structural connectivity mirrors that of the TRN. We have analyzed linearly stable states of the model near
instability boundaries representing precursive states to full seizure
and oscillations characteristic of Parkinson's disease.
%
%
The main results of the paper are as follows:

(i) A simplified model of the CTBG system was developed that enables parallel analysis of PD and epilepsy in a tractable framework. The model successfully reproduces PD rhythms and levels of activity present in a previous, more detailed, model of the CTBG system~\citep{Albada09a,Albada09b}.

(ii) By parametrizing key loops, the stability of the CTBG system was represented in a dimensionally reduced state space~\citep{Robinson02}. Resonances of these key loops were shown to define the spectra of population activity.

(iii) Delta/theta activity around $4$~Hz was demonstrated in the BG which is consistent with oscillations typically observed in Parkinson's disease~\citep{Brown01,Timmermann03,Wang05}.

(iv) The BG were also shown to support beta oscillations around $20$~Hz.
This activity is consistent with an experimentally observed coherence peak between BG that constitute the indirect pathway~\citep{Brown01,Brown05,Levy02}.

(v) Spindle-generating sleep states, present in a previous CT model~\citep{Robinson02}, were found in the CTBG system. The results suggested similar sleep spindle states for healthy and PD patients because the spindle frequency is generated by a dominant intrathalamic loop resonance.

(vi) The simplified CTBG framework predicted a reduced likelihood of tonic-clonic (grand mal) seizures in parkinsonian states and agrees with the results of experiment~\citep{Velivskova96,Dybdal00}.

(vii) The model also predicts a greater likelihood of absence seizure (petit mal) in PD wake states due to an increased proximity to the delta/theta instability boundary, and could explain the observed increased occurrence of absence seizures after dopaminergic depletion~\citep{Danober98,Deransart98,Deransart00,De00,Mid01}.

(viii) Using the CTBG model we also explored the role of the BG in modulating states near absence seizure. The system is placed in close proximity to the delta/theta instability boundary where enhanced 3~Hz activity results from a dominant corticothalamic loop resonance.
These oscillations were shown to be suppressed by increasing the strength of afferent and efferent BG connections to corticothalamic populations. This was independent of which connection parameter was targeted.
Instead, the suppressive effect is determined by a weakened dominance of the cortico-TRN-SRN neural circuit responsible for the 3~Hz oscillations, relative to the cortico-BG-SRN and the more direct cortico-SRN neural circuits.

(ix) Strengthening cortico-BG and BG-SRN connections moved states of the model away from the alpha instability boundary. The boundary defines a transition to tonic-clonic seizure states where the alpha resonance, producing $\sim$$10$~Hz activity, is no longer damped.
This agrees with the efficacy of tonic-clonic
seizure treatments targeting neural circuits involving the BG.

(x) Finally, the CTBG model showed that strong afferent and efferent TRN connections, such as those in sleep states, have a suppressive effect on parkinsonian activity of the BG. This result is consistent with circadian fluctuations of motor-symptoms typical in PD~\citep{Marsden82,Fahn82} and the absence of parkinsonian tremor and rigidity during sleep~\citep{Broughton67,Stern68}.

%



%
Turning to wider implications of the work, we note that the presence of $4$~Hz delta/theta oscillations in the activity of the simplified BG population implies synchronization within and between intrinsic BG nuclei.
Since the oscillations result from a resonance in the pathway formed between the BG, thalamus and cortex, the activity of nuclei comprising the dominant pathway through the BG must be coherent at the delta/theta resonance frequency.
This result is consistent with the significant theta band coherence observed by \cite{Brown01} between STN and GPi nuclei, and by \cite{Levy02} between GPi and GPe nuclei. 
%
%
It also provides an explanation for the coherence of LFP oscillations in the ventro-intermediate nucleus of the thalamus~(VIM) with tremor recordings~\citep{Tass10}.

\cite{Robinson02} defined a CT model that, in addition to other phenomena, accounted for the $\sim$$14$~Hz spindle oscillations observed during sleep stage 2. In this model, the spindle frequency results from a resonance in the intrathalamic loop formed between the TRN and SRN.
An analogous circuit in the CTBG model, formed between the BG and SRN, is able to sustain $\sim$$20$~Hz beta band oscillations.
These $\sim$$20$~Hz oscillations are predicted to be coherent between nuclei comprising the dominant pathway within the BG.
This agrees with experimental studies of PD which showed coherent beta band activity at $\sim$$20$~Hz between STN and GPi nuclei, and between GPi and GPe nuclei~\citep{Brown01,Brown05,Levy02,Kuhn05}.
It could also explain the direct coupling of beta band LFP oscillations in the STN to fluctuations in synchronized activity of local neurons~\citep{Kuhn05} which has been suggested to play a key role in the pathophysiology of PD~\citep{Levy02dep}.

The overall effect of dopamine loss in the CTBG model is to move the system towards the delta/theta and sigma instability boundaries. In the vicinity of these boundaries the delta/theta and sigma resonances are weakly damped.
The BG are an integral part of the dominant loops facilitating the delta/theta and sigma resonances in PD.
Thus, the model predictions are consistent with the enhanced theta and beta band oscillations which are a predominant feature of parkinsonian BG activity.

%
%

%
Overall, our simplified CTBG model demonstrates that the role of the BG in generating parkinsonian states may be analogous to the role of TRN in generating epileptic states and lays the foundation for further exploration of PD and epilepsy in a single unified framework.

The analysis presented in this work is a linear approximation
to the full nonlinear set of partial delayed differential equations
representing the corticothalamic-basal ganglia system. This limits
the results to parameter regimes where the system is dominated
by linear dynamics. Future work will look to explore the full nonlinear case. The model also uses effective values for the parameters describing GABAergic inhibitory synaptic responses. Both fast
and slow kinetics of the GABAergic synapse appear to play an important role in the generation of epileptiform activity and thus this
represents as a limitation to the work.

\section{Acknowledgments}
The authors thank J. C. Pang and S. Assadzadeh for stimulating discussions.
This work was supported by the Australian Research Council Center of Excellence for Integrative Brain Function under ARC grant CE140100007 and by Australian Research Council Laureate Fellowship Grant FL140100025.

\bibliographystyle{elsarticle-harv} 
\bibliography{Muller2016}

\end{document}